\documentclass[3p,times]{elsarticle}

\usepackage{ecrc}

\volume{00}

\firstpage{1}

\journalname{Nuclear Instruments and Methods in Physics Research Section A}

\runauth{F. Mantegazzini et al.}

\jid{NIM A}

\usepackage{amssymb}

\usepackage[utf8]{inputenc}
\usepackage{upgreek}
\usepackage{graphicx}
\usepackage{amsmath}
\usepackage{makecell}
\usepackage{subcaption}
\usepackage[shortlabels]{enumitem}
\usepackage{multicol}
\usepackage{xcolor}
\usepackage{hyperref}

\usepackage{upquote} 

\usepackage[figuresright]{rotating}

\begin{document}

\begin{frontmatter}

\dochead{}

\title{Metallic magnetic calorimeter arrays \\ for the first phase of the ECHo experiment}

\author[1]{F. Mantegazzini}
\author[1]{A. Barth}
\author[4]{H. Dorrer}
\author[4,5,6]{Ch. E. Düllmann}
\author[1]{C. Enss}
\author[1]{A. Fleischmann}
\author[1]{R. Hammann}
\author[2,1]{S. Kempf}
\author[3,4,5,6]{T. Kieck}
\author[1]{N. Kovac}
\author[1]{C. Velte}
\author[2,1]{M. Wegner}
\author[3]{K. Wendt}
\author[1]{T. Wickenhäuser}
\author[1]{L. Gastaldo}

\affiliation[1]{organization={Kirchhoff Institute for Physics, Heidelberg University},
            addressline={INF 227}, 
            city={Heidelberg},
            postcode={D-69120}, 
            country={Germany}}
            
\affiliation[2]{organization={Institute of Micro- and Nanoelectronic Systems, Karlsruhe Institute of Technology},
            addressline={Hertzstrasse 16}, 
            city={Karlsruhe},
            postcode={D-76187}, 
            country={Germany}}
            
\affiliation[3]{organization={Institute of Physics, Johannes Gutenberg University Mainz},
            addressline={Staudingerweg 7}, 
            city={Mainz},
            postcode={D-55128}, 
            country={Germany}}

\affiliation[4]{organization={Department of Chemistry - TRIGA Site, Johannes Gutenberg University Mainz},
            addressline={Fritz-Stra\ss mann-Weg 2}, 
            city={Mainz},
            postcode={D-55128}, 
            country={Germany}}
            
\affiliation[5]{organization={Helmholtz Institute Mainz},
            addressline={Staudingerweg 18}, 
            city={Mainz},
            postcode={D-55128}, 
            country={Germany}}        
            
\affiliation[6]{organization={GSI Helmholtzzentrum für Schwerionenforschung GmbH},
            addressline={Planckstra\ss e 1}, 
            city={Darmstadt},
            postcode={D-64291}, 
            country={Germany}}

\begin{abstract}

\noindent The ECHo experiment has been designed for the determination of the effective electron neutrino mass by means of the analysis of the end-point region of the $^{163}$Ho electron capture spectrum. Metallic magnetic calorimeters enclosing $^{163}$Ho are used for the high energy resolution calorimetric measurement of the $^{163}$Ho spectrum.
For the first phase of the experiment, ECHo-1k, a 72-pixel MMC array has been developed. The single-pixel design has been optimised to reach 100\% stopping power for the radiation emitted in the $^{163}$Ho electron capture process (besides the electron neutrino) and an energy resolution $\mathrm{\Delta}E_{\mathrm{FWHM}} < 10 \, \mathrm{eV}$.
We describe the design of the ECHo-1k detector chip, the fabrication steps and the characterisation at room temperature, at $\mathrm{4 \, K}$ and at the final operation temperatures.
In particular, a detailed analysis of the results from these tests allowed to define a quality check protocol based on parameters measurable at room temperature.
We discuss the performance achieved with the two ECHo-1k detector chips – the first one with $^{163}$Ho implanted in gold and the second one with $^{163}$Ho implanted in silver – which have been used for the high statistics measurement of the ECHo-1k experiment. An average activity per pixel of $0.81 \, \mathrm{Bq}$ and $0.71 \, \mathrm{Bq}$ and an average energy resolution of $6.07 \, \mathrm{eV}$ FWHM and $5.55 \, \mathrm{eV}$ FWHM have been achieved with these two detectors, fulfilling the requirements for the first phase of the ECHo experiment.

\end{abstract}


\end{frontmatter}

\section{Introduction}

The determination of neutrino masses is a fundamental goal of the current research in particle physics. Direct and model-independent methods to measure this quantity are based on the study of beta decay and electron capture (EC) processes \cite{NOW2016} \cite{Drexlin_Review}.
In a beta decay, the parent atom decays to the daughter atom emitting an electron (positron) and an electron antineutrino (electron neutrino).
In an EC process, on the other hand, an electron from an inner atomic shell is captured by the nucleus and an electron neutrino is emitted. The daughter atom is left in an excited state characterised by the vacancy left by the captured electron. The excitation energy is then released through the emission of X-rays and/or electrons. 
In the case of beta decay, the kinetic energy of the emitted electrons is measured, while in the case of EC the de-excitation energy of the daughter atom is the experimentally accessible quantity.
In both cases, the information on a finite neutrino mass derives from the neutrino phase space term. In particular, the largest effect can be seen in the region just below the end-point of the spectrum.

The current best limit on the effective electron antineutrino mass is $\mathrm{0.8 \, eV}$ 90\% C.L. and it was recently achieved by the KATRIN collaboration \cite{KATRIN_PRL_2019} \cite{KATRIN_limit_2021}.
For the effective electron neutrino mass, however, the best limit is $\mathrm{150 \, eV}$ 90\% C.L. and it was obtained by the ECHo collaboration with a calorimetric measurement of the $^{163}$Ho EC spectrum \cite{ECHo_spectrum_2019}. This result improved the limit of 225 eV 95\% C.L. achieved in 1987 by analysing the Internal Bremsstrahlung in Electron Capture (IBEC) spectrum of $^{163}$Ho \cite{Springer_IBEC}.

Experiments based on the $^{163}$Ho EC process, like ECHo \cite{ECHoGeneral} and Holmes \cite{HOLMES}, plan to achieve sensitivity levels similar to the ones currently obtained with $^3$H-based approaches.
The strategy to achieve this goal is based on the analysis of the calorimetrically measured EC spectrum of $^{163}$Ho, as proposed by De Rujula and Lusignoli \cite{DeRujula_Lusignoli}. To perform such a measurement, the $^{163}$Ho source needs to be enclosed in high energy resolution detectors, so that all the energy emitted in the decay - besides the neutrino energy - is measured. \\
The resulting energy spectrum is characterised by several resonances, one for each excited state in which the daughter atom can be left, and by continuous tails which derive from electrons and photons with a continuous energy spectrum \cite{Brass_HoSpectrum_2018} \cite{Brass_HoSpectrum_2020}.
The maximum energy for the excitation of the $^{163}$Dy atom is given by the difference between the mass of the parent atom and the one of the daughter atom $Q_{\mathrm{EC}} = 2.833 \pm 0.030_{\mathrm{stat}} \pm 0.015_{\mathrm{sys}} \, \mathrm{keV}$ - as determined via Penning Trap Mass Spectrometry \cite{Eliseev_Q-value}.
This value perfectly agrees with the one derived from the analysis of the calorimetrically measured $^{163}$Ho spectrum: $Q_{\mathrm{EC}} = 2.838 \pm 0.014 \, \mathrm{keV}$ \cite{ECHo_spectrum_2019}. \\
Because of energy conservation, no K and L captures are possible and the first electrons which can be captured belongs to the 3s-shell (M capture).
The nuclide $^{163}$Ho is the best candidate for the determination of the effective electron neutrino mass because of the very low Q-value, which is relatively close to the energy of the MI resonance. The end-point part of the spectrum is dominated by the right tail of the MI resonance, increasing the fraction of events in the region of interest (ROI), i.e.~in the last few eV before the end-point.

In the ECHo experiment, metallic magnetic calorimeters (MMCs) loaded with $^{163}$Ho are employed \cite{MMC_AIP} \cite{Gastaldo_NIMA}. For the first phase, ECHo-1k \cite{ECHoGeneral}, MMC arrays hosting 72 pixels with dc-SQUID read-out are used. The aim of this phase is to reach about 10$^8$ events in order to obtain a sensitivity below $20 \, \mathrm{eV}$ on the effective electron neutrino mass. \\
For ECHo-1k, dedicated detector chips have been developed and characterised, showing an optimal performance that fully matches the requirements of this phase. \\
We discuss the design, fabrication and quality control of the ECHo-1k detector chip. In particular, we report the full characterisation of the two ECHo-1k chips used for the high statistics measurement at the end of the ECHo-1k experimental phase.
\section{Metallic magnetic calorimeters}
\label{SEC:MMCs}

The detectors used in the ECHo experiment are metallic magnetic calorimeters (MMCs) operated at millikelvin temperature. These calorimeters consist of a particle absorber, with dimensions optimised for the particles to be detected, placed in good thermal contact with a paramagnetic thin film sitting in a constant magnetic field, which acts as a sensitive thermometer \cite{Fle2005}. This temperature sensor consists of Er$^{3+}$ ions diluted in silver with a concentration of few hundred ppm and is weakly connected to a thermal bath at constant temperature of $\mathrm{20 \, mK}$ or below. When a particle interacts in the absorber, a certain energy $\Delta E$ is deposited, causing an increase of temperature $\Delta T$. Consequently, the magnetisation of the paramagnetic sensor decreases. This change of magnetisation $\Delta M$ is translated into a change of flux $\Delta \Phi$ in a superconducting pick-up coil connected in parallel to the input coil of a direct current Superconducting Quantum Interference Device (dc-SQUID). 
The corresponding change of flux in the dc-SQUID leads to a change of voltage which can be linearised using a feedback loop \cite{Drung}.
In summary, the measured change of voltage $\Delta V$ is proportional to the energy deposited in the MMC $\Delta E$, according to:

\begin{equation}\label{MMC_signal}
\Delta V \propto \Delta \Phi \propto \frac{\partial M}{\partial T} \Delta T \simeq \frac{\partial M}{\partial T} \frac{\Delta E}{C}
\end{equation}

where $C$ is the total heat capacity of the detector. \\

MMC detectors developed for soft X-ray spectroscopy have shown excellent energy resolution, reaching 1.6 eV full width half maximum (FWHM) at 6 keV \cite{Kem2018}. The signal rise-time is about 90 ns and limited by the spin-electron interaction \cite{Kem2018}. These detectors show a good linear behaviour with only small quadratic deviation leading to a non linear contribution of about 1\% at $6 \, \mathrm{keV}$ for detectors optimised for X-ray spectroscopy \cite{Kem2018}.

To achieve the goal of the first ECHo phase, MMCs had to be optimised to fulfil the following requirements:

\begin{enumerate}

    \item 4$\uppi$ geometry and high quantum efficiency to guarantee a calorimetric measurement of all the energy realised in the decay $\rightarrow$ $^{163}$Ho source between two $\mathrm{5 \, \upmu m}$ thick gold absorbers;
    
    \item optimised values of $^{163}$Ho activity, not to exceed defined values of unresolved pile-up and heat capacity of the absorber $\rightarrow$ $\mathrm{A \leq 10 \, Bq}$;

    \item excellent energy resolution to avoid smearing of the $^{163}$Ho spectrum $\rightarrow$ $\mathrm{\Delta E_{FWHM} \leq 10 \, eV}$ for energies below $\mathrm{10 \, keV}$;
    
    \item fast signal rise to minimise the chance that two events occur within the rise part of the signa, therefore being not distinguishable, leading to unresolved pile-up $\rightarrow$ $\mathrm{\tau} < 1 \, \upmu$s;
    
    \item reliable detector calibration for a correct energy scale of the $^{163}$Ho spectrum and to precisely define the ROI where to search for modifications of the spectral shape due to the finite neutrino mass.

\end{enumerate}

Gold is the material of choice for the absorbers due to high $Z$, chemical stability, good thermal conductance as well as possibility to reliable fabricate gold absorbers using electroplating and to reach high RRR values for internal thermalisation. 
The distinctive feature of the detectors for ECHo is the enclosing of $^{163}$Ho in the absorber. 
The absorber needs to be designed to allow for highly reliable implantation of $^{163}$Ho atoms in the required amount.
It has already been demonstrated that it is possible to ion-implant $^{163}$Ho atoms into MMCs absorbers. The very preliminary tests employing MMCs for the calorimetric measurement of the $^{163}$Ho spectrum \cite{Gastaldo_NIMA}, for which the $^{163}$Ho enclosure was obtained via ion-implantation at the ISOLDE facility (CERN), have demonstrated the suitability of MMCs as well as of the implantation technique \cite{Ranitzsch_PRL}\cite{Ranitzsch_JLTP}.
\section{ECHo-1k detector design}
\label{SEC:detector_design}

\begin{figure}[h!]
    \centering
    \begin{subfigure}[b]{0.3\textwidth}
        \centering
        \includegraphics[width=\linewidth]{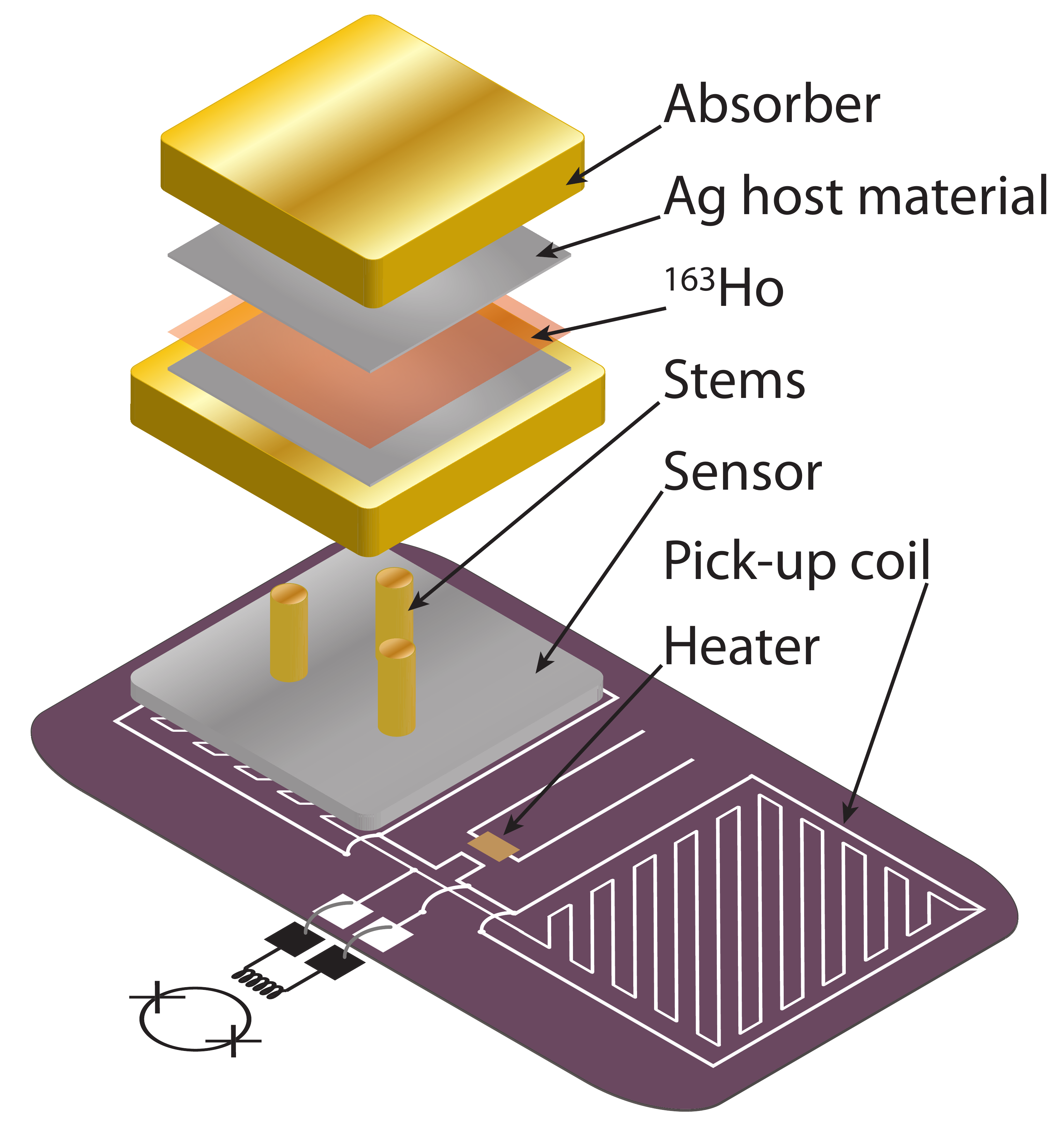}
        \caption{} \label{SUBFIG:blowup_ch}
    \end{subfigure}
    \hfill
    \begin{subfigure}[b]{0.5\textwidth} 
        \centering
        \includegraphics[width=\linewidth]{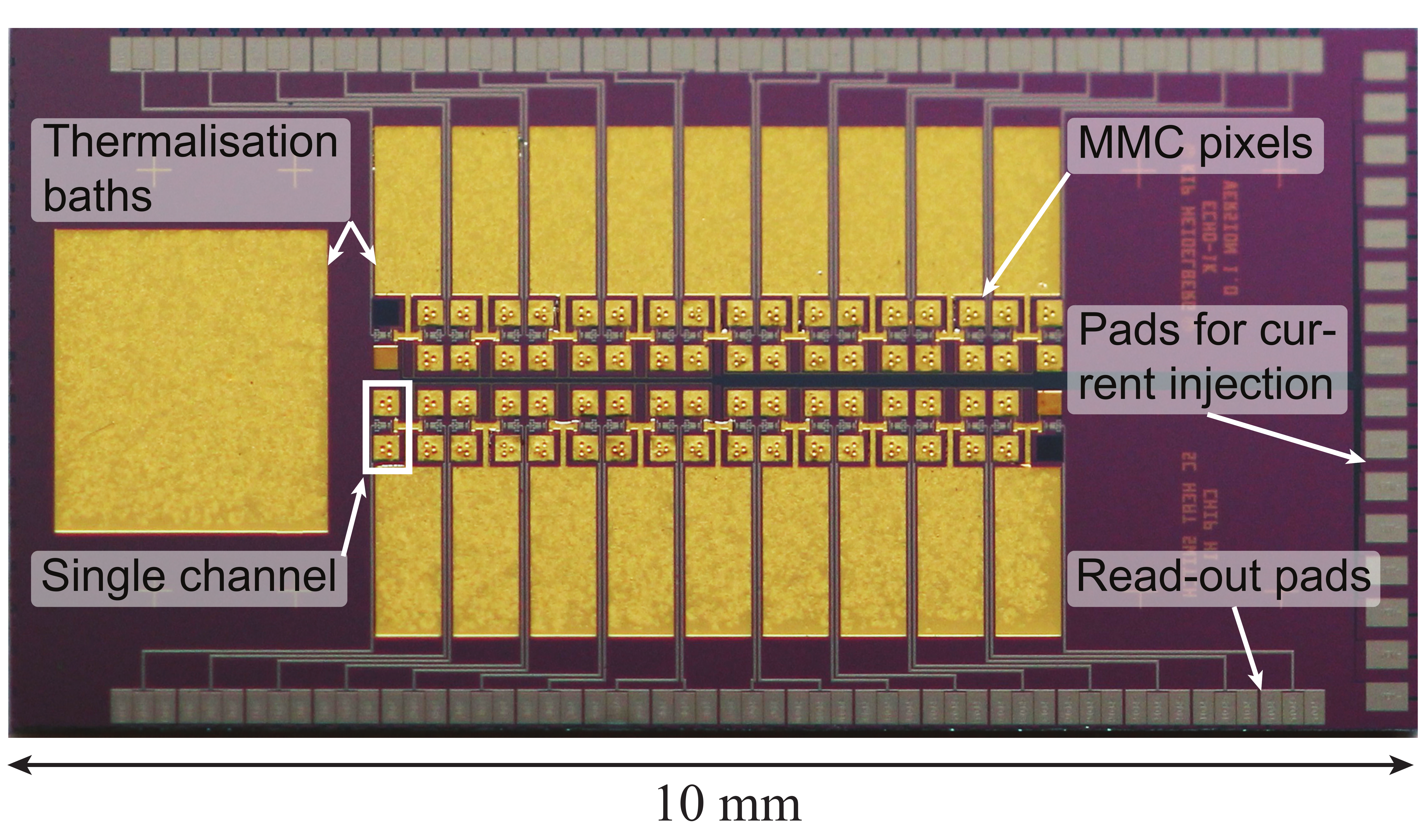}
        \caption{} \label{SUBFIG:ECHo-1k_chip}
    \end{subfigure}

    \caption{\textbf{a)} Schematic layout of a single detector channel comprising two pixels. \textbf{b)} ECHo-1k array chip with 36 MMC channels. Each single detector channel consist of two pixels. The gold thermalisation baths, the read-out pads and the pads for current injection are highlighted.}

    \label{FIG:ECHo-1k}
\end{figure}

For the first phase of the ECHo experiment, ECHo-1k, a dedicated MMC array design has been developed. It consists of an array of 36 MMCs featuring a planar detector geometry with meander-shaped pick-up coils \cite{MMC_AIP}. Each MMC consists of two detector pixels. For 34 of these detectors both pixels are equipped with sensor and absorber. The input coil of a dc-SQUID is connected in parallel to the pick-up coils forming a first order gradiometer.
Figure \ref{SUBFIG:blowup_ch} shows an exploded-view drawing of a detector channel of the ECHo-1k chip and figure \ref{SUBFIG:ECHo-1k_chip} shows a microscope photograph of the entire chip. 
The chip has a size of $10 \, \mathrm{mm} \times 5 \, \mathrm{mm}$ and is divided into four quarters, each including nine detectors, i.e.~18 detector pixels. The geometry and the size of the single detector pixel are described in section \ref{SUBSEC:ECHo-1k_fabrication}.

Two channels, placed at two opposing corners of the array, are dedicated to temperature monitoring. They are characterised by a non-gradiometric layout, having only one of the two pixels equipped with a temperature sensor. They are therefore sensitive to temperature fluctuations of the substrate. The temperature information obtained with these channels is correlated to the amplitude of the detector signals acquired with the other channels of the array and it is used to perform a temperature correction of the signal amplitudes  against temperature drifts.  The temperature monitoring channels are not loaded with $^{163}$Ho.

Two channels at the other two opposing corners have not been implanted and they can be used for diagnostics and characterisation studies, in particular for comparing the performances of the two pixels belonging to the same detector and, in general, for testing the performances of non-implanted MMCs. 

The remaining 32 channels in the central area of the chip are used for the measu\-re\-ment of the $^{163}$Ho spectrum and for \textit{in situ} background measurements. For this the implantation mask is designed so that:
\begin{itemize}
    \item 25 channels have both pixels implanted enclosing $^{163}$Ho;
    \item the remaining seven channels have only one pixel loaded with $^{163}$Ho, so that the empty pixel allows for \textit{in situ} background measurements.
\end{itemize}

The non-implanted pixels dedicated to background studies are crucial to constantly monitor the background level and they allow for a direct comparison of the background data with the simulated background models.
The calibration of these non-implanted pixels can be achieved exploiting the calibration curve determined for the second pixel implanted with $^{163}$Ho belonging to the same channel.

\begin{figure}[h!] 
    \centering
    \includegraphics[width=0.4\textwidth]{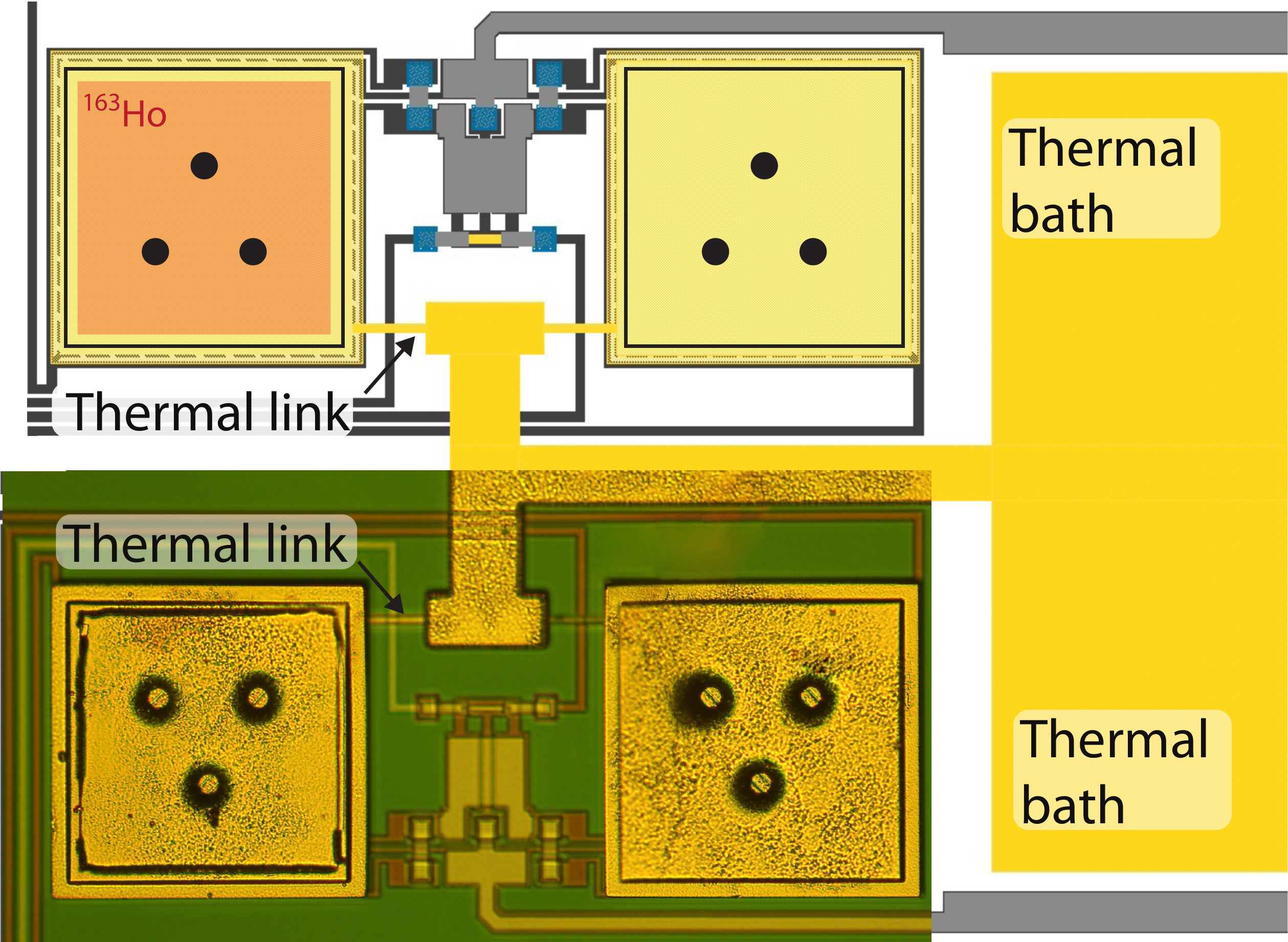}
    \caption{Design (pixel pair at the top) and microscope picture (pixel pair at the bottom) of the thermalisation concept in the ECHo-1k design. Four fully fabricated ECHo-1k pixels are shown. The two left pixels are implanted with $^{163}$Ho. The weak thermal link and the thermal bath are indicated.} 
    \label{FIG:thermal_link}
\end{figure}

The connection of each pixel to the thermal bath is shown in figure \ref{FIG:thermal_link} and can be described by three parts connected in series, namely a weak thermal link, a T-shaped gold area and a large thermalisation bath.
The weak thermal link which consists of a gold film with an area of $42.5 \, \mathrm{\upmu m} \times 5.0 \, \mathrm{\upmu m}$ and a thickness of $0.3 \, \mathrm{\upmu m}$ and connects the sensor to the T-shaped gold area.
The T-shaped gold area is in turn connected on chip to the thermalisation bath with an area of $0.56 \, \mathrm{mm^2}$ and a thickness of $5 \, \mathrm{\upmu m}$. The heat capacity of the thermal bath is $4.0 \, \mathrm{pJ/K}$ at a temperature $T = 20 \, \mathrm{mK}$.
Each thermal bath can be connected via gold wire-bonds to the next one and finally to the larger gold thermalisation bath (figure \ref{SUBFIG:ECHo-1k_chip}). The large thermalisation bath is intended to be connected via gold wire-bonds to the experimental copper holder to guarantee optimal thermalisation.

\subsection{Detector fabrication}
\label{SUBSEC:ECHo-1k_fabrication}

The ECHo-1k detectors have been fabricated on 3\textquotesingle\textquotesingle$\,\,$silicon wafers in the cleanroom facility of the Kirchhoff-Institute for Physics, Heidelberg University. Each wafers hosts 64 chips. The fabrication is divided into 14 steps (including the $^{163}$Ho implantation procedure) which are summarised in table \ref{TAB:fabrication}.

\vspace{0.2cm}

\begin{table}[h!]
\centering
\begin{tabular}{ |p{0.4cm}|p{4.1cm}|p{1.3cm}|p{1.6cm}|p{4.2cm}|  }
 \hline
 \multicolumn{5}{|c|}{Fabrication steps} \\
 \hline
 \hline 
 \# & Photo-lithographic mask & Material & Thickness & Deposition technique\\
 \hline
 1 & \makecell[l]{Pick-up coils, \\ SQUID lines (layer 1)}  & Nb & 300 nm & Sputtering + etching\\
 2 & First isolation layer  & \makecell[l]{$\mathrm{Nb_{2}O_{5}}$ \\ $\mathrm{SiO_{2}}$} & \makecell[l]{- \\ 175 nm} & \makecell[l]{Anodisation \\ Sputtering + lift-off}\\
 3 & Second isolation layer  & $\mathrm{SiO_{2}}$ & 175 nm & Sputtering + lift-off\\
 4 & Heaters  & AuPd & 150 nm & Sputtering + lift-off\\
 5 & SQUID lines (layer 2)  & Nb & 600 nm & Sputtering + lift-off\\
 6 & Sensor  & AgEr & 1.3 $\mathrm{\upmu m}$ & Sputtering + lift-off\\
 7 & Thermal link  & Au & 300 nm & Sputtering + lift-off\\
 8 & Stems  & Au & 200 nm & Sputtering\\
 9 & First absorber \& thermal bath & Au & 5 $\mathrm{\upmu m}$ & Electroplating + lift-off\\
 10 & $^{163}$Ho host material & Ag & 150 nm & Sputtering\\
 11 & $^{163}$Ho implantation & $^{163}$Ho & - & Ion-implantation\\
 12 & $^{163}$Ho host material & Ag & 150 nm & Sputtering + lift-off\\
 13 & Second absorber & Au & 5 $\mathrm{\upmu m}$ & Sputtering + lift-off\\
 \hline
\end{tabular}
\caption{Microfabrication steps of the ECHo-1k detector chip.}
\label{TAB:fabrication}
\end{table}

The single MMC detector comprising two pixels is depicted in figure \ref{SUBFIG:blowup_ch}. The corresponding fabrication process starts with the production of a niobium superconducting meander-shaped pick-up coil, which is used to detect the magnetic response of the sensor and to create the constant bias magnetic field by means of a persistent current. For this purpose, a 300 nm thick niobium layer is sputter deposited on the full substrate, then photoresist is structured on top covering the areas that are part of the design and finally the rest of the material is removed via reactive-ion etching\footnote{Etching process performed with ICP Cryogenic Etch System SI 500 C.}. 
The meander-shaped pick-up coil is formed by 39 stripes with a width of 3$\mathrm{\,\upmu m}$ and a pitch of 6$\mathrm{\,\upmu m}$. The total surface of a single pick-up coil is $\mathrm{172 \, \upmu m^2}$.
The two pick-up coils belonging to one detector channel are connected to the bonding pads at the rim of the chip (as shown in figure \ref{SUBFIG:ECHo-1k_chip}). 

For isolation purposes, the niobium structures are anodised and two $\mathrm{SiO_{2}}$ layers are deposited on top (as reported in Table \ref{TAB:fabrication}). For this, a lift-off technique is employed: photoresist is structured on the substrate leaving open the areas which need to be isolated, $\mathrm{SiO_{2}}$ is then sputter deposited on top and finally the resist is lifted-off. 

The next layer contains normal conducting heater elements which are used for the injection of the persistent current in the superconducting niobium pick-up coils (as explained in detail in the subsection \ref{SUBSEC:MMC_operation}). The heaters consist of Au:Pd film with a nominal thickness of $150 \, \mathrm{nm}$. After the sputter deposition of a thin niobium sticking layer, the heaters are fabricated via a sputtering and lift-off process. Finally, a thin niobium layer is sputter deposited as protection layer.

After that, a second niobium layer is deposited to complete the strip-lines that connects the pick-up coils to the bonding pads.
A strip-line consists of two overlapping connection lines separated by an isolation layer in order to minimise the parasitic inductance.

On top of the pick-up coil, after the sputter deposition of a thin niobium sticking layer, the Ag:Er sensor layer with an erbium concentration of about $445 \, \mathrm{ppm}$ and with an area of $\mathrm{168\,\upmu m} \times \mathrm{168\,\upmu m}$ is sputter deposited and lifted-off. The thickness of the sensor is designed to be $1.3 \, \mathrm{\upmu m}$, but a gradient of few percent over the wafer can occur.

In the next layer, the gold thermal link with a thickness of $300 \, \mathrm{nm}$ that connects the sensor to the thermal bath is fabricated via sputtering deposition, after a thin niobium sticking layer.

On top of the sensor, a photoresist layer defining three gold stems with a diameter of 10$\mathrm{\,\upmu m}$ for the support of the first absorber layer is structured and a $200 \, \mathrm{nm}$ tick gold layer is sputtered. No lift-off is performed at this point. \\
A second photoresist layer, defining the absorber area of $\mathrm{180 \,\upmu m} \times \mathrm{180 \,\upmu m}$, is structured on top of the previous one. 
Finally, the stems and the absorber are fabricated simultaneously using the TECHNI-GOLD 25 ES gold plating solution\footnote{Produced by Technic Inc., Cranston, Rhode Island.} and a current per area of $\mathrm{1 \, mA/cm^2}$. The absorber thickness is designed to be $5 \, \mathrm{\upmu m}$. \\
The purpose of the stems is to minimise the contact area between sensor and absorber. This prevents athermal phonons from travelling through the sensor and releasing the energy in the substrate. This energy would therefore not be detected leading to a low energy asymmetry in the detector response \cite{Hassel_JLTP}. 

\subsection{$^{163}$Ho implantation} \label{SUBSEC:implantation}

\begin{figure}[h!] 
    \centering
    \includegraphics[width=0.6\textwidth]{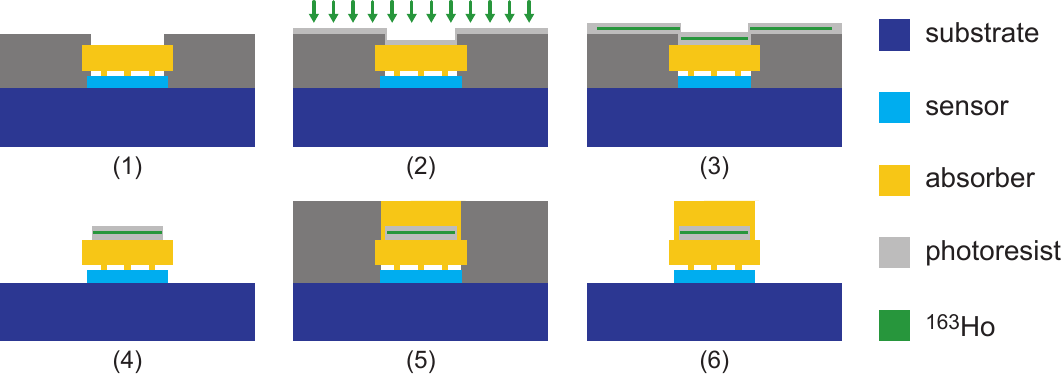}
    \caption{The microfabrication steps for implanting $^{163}$Ho in the ECHo-1k MMC pixels (corresponding to the layers 11, 12, 13 in \ref{TAB:fabrication}) are schematically depicted. The photoresist is shown in dark grey; the host-material as well as the sensor are shown in light grey; the $^{163}$Ho source is shown in green; the gold absorber is shown in yellow.} 
    \label{FIG:implantation_steps}
\end{figure}

The first attempt of $^{163}$Ho on-line ion implantation in MMC detectors was performed at ISOLDE with successful results regarding the detector performance, but radioactive contamination could not be avoided \cite{Gastaldo_NIMA}. 
Off-line implantation of resonantly laser-ionised and mass-separated ions, starting from a chemically purified holmium sample has been chosen to prevent the presence of radioactive nuclides other than $^{163}$Ho in the implantation beam. The $^{163}$Ho has been produced by intense
neutron irradiation of an Er target enriched in $^{162}$Er in the high flux research reactor at Institute Laue-Langevin, Grenoble, France. After irradiation, the holmium was chemically separated from the target material \cite{Dorrer}. The holmium samples still contain all holmium isotopes present in the target after irradiation: the desired $^{163}$Ho, stable $^{165}$Ho, and traces of $^{166\mathrm{m}}$Ho ($^{166\mathrm{m}}$Ho:$^{163}$Ho $\approx 10^{-4}$). The latter undergoes beta decay with a half-life of about 1200 years and a Q-value of $1860 \, \mathrm{keV}$. Its presence in the detector will contribute to undesired background, necessitating its separation \cite{Dorrer}. Both undesired isotopes, $^{165}$Ho and $^{166\mathrm{m}}$Ho, can be mass separated from $^{163}$Ho. The mass separation and ion implantation are performed at the RISIKO facility at Johannes Gutenberg University Mainz \cite{RISIKO}. The RISIKO system employs efficient ionisation in a laser ion source \cite{Kieck_laser_ion_source}, acceleration of extracted ions to $30 \, \mathrm{keV}$ and a magnet for mass separation. The separated beam undergoes post-focusing and is then directly implanted into the MMC detectors mounted at the RISIKO second focal plane. The overall efficiency is about 70\%. The $^{166\mathrm{m}}$Ho is suppressed to $\leq 1.4(6) \times 10^{-5}$ of its initial fraction, relative to $^{163}$Ho. Considering this suppression value together with the initial fraction of $^{166\mathrm{m}}$Ho present in the sample, we expect a ratio of $^{166\mathrm{m}}$Ho:$^{163}$Ho $\approx 10^{-9}$ in the detector \cite{RISIKO}.
The microfabrication steps which are necessary to enclose $^{163}$Ho in the MMC detectors are depicted in figure \ref{FIG:implantation_steps}.

\begin{figure}[h!] 
    \centering
    \includegraphics[width=0.42\textwidth]{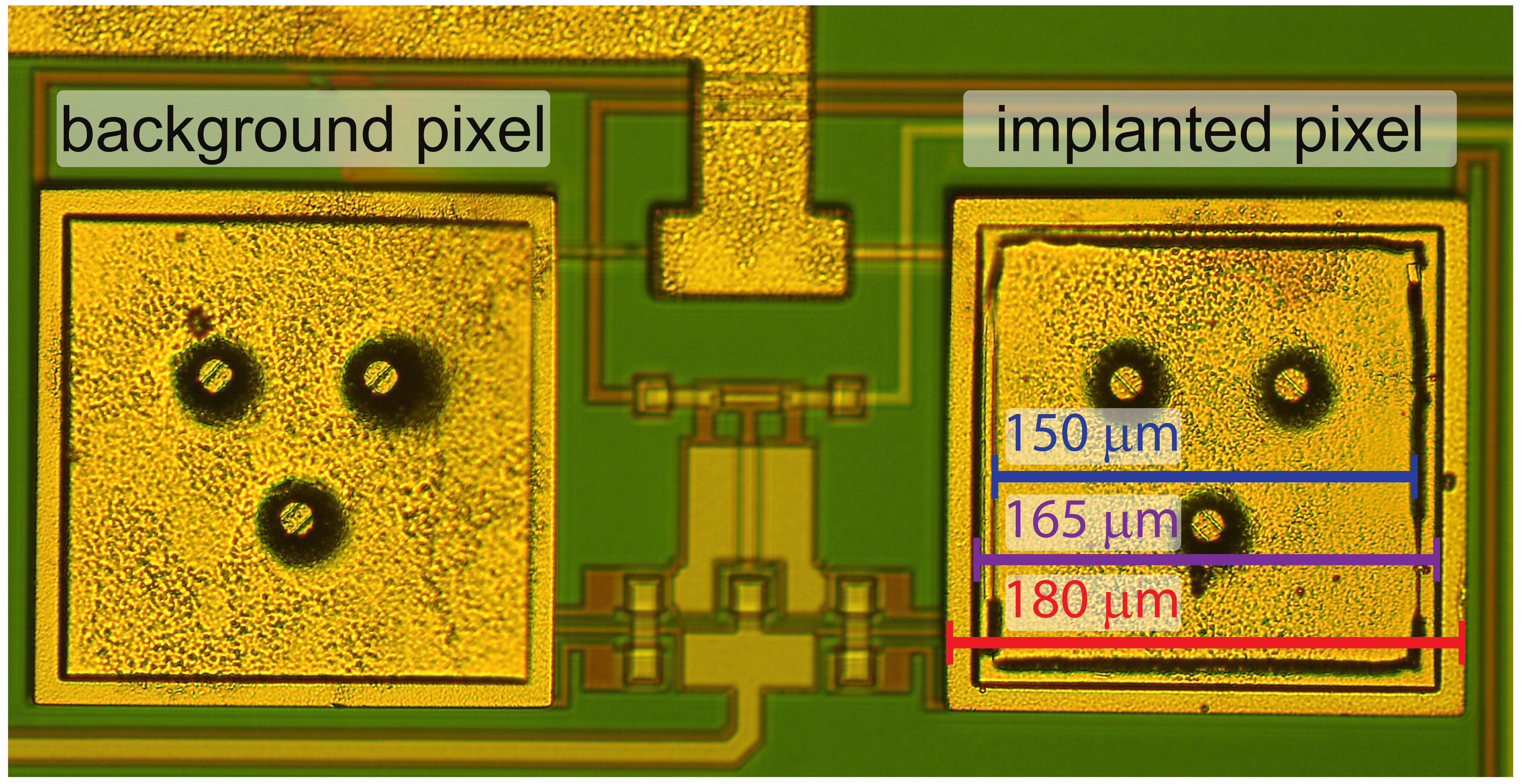}
    \caption{A microscope photograph of the two pixels of a post-processed detector is shown. The pixel on the left is a background pixel, i.e. with no $^{163}$Ho implanted, while the pixel on the right is loaded with $^{163}$Ho. The implantation layer is marked in blue, while the top (bottom) absorber is marked in purple (red).} 
    \label{FIG:pixel_postproc}
\end{figure}

First, a photoresist mask is applied on a single ECHo-1k chip leaving open only areas of $\mathrm{(150 \, \upmu m)^2}$ on top of each absorber to be implanted. 
A thin layer of typically $100 \, \mathrm{nm}$ of host material of choice is then sputter deposited on the full substrate and further exploited to contact to ground and to avoid charging of the sample during implantation. Different host materials - gold, silver and aluminium - have been considered and tested, as explained in the following. The $^{163}$Ho implantation depth depends on the selected material and on the ion energy. Assuming the operational ion energy of 30 keV used at RISIKO, a mean longitudinal implantation depth of few nanometers is expected in silver and gold \cite{implantation_sim}. 
As next step, $^{163}$Ho is ion-implanted into the host material. To protect the implanted area, a second thin layer of typically $100 \, \mathrm{nm}$ of the host material is deposited additionally over the sample within a few days after the implantation process. 
Finally, the chip undergoes the post-processing phase. The photoresist mask used for implantation is lifted-off and a second photoresist mask is prepared for the structuring of the second part of the absorber. For this a $\mathrm{5 \, \upmu m}$ gold film is structured on top of each pixel (including the background pixels) over an area of $\mathrm{(165 \, \upmu m)^2}$.
Figure \ref{FIG:implantation_steps} shows the post-process steps for $^{163}$Ho implantation and structuring of the top absorber layer.
Figure \ref{FIG:pixel_postproc} shows a microscope image of the final layouts of an implanted MMC pixel as well as a background not-implanted pixel. The implanted pixel can be distinguished by the presence of three square structures: the innermost one (marked in blue) corresponds to the implantation area, the outermost one (marked in red) to the edge of the bottom absorber layer and the one in the middle (marked in purple) to the top absorber layer. On the left side, the background pixel does not show the square structure corresponding to the implantation mask, but the top and bottom absorber layers are visible.

Different $^{163}$Ho host materials have been considered and tested during the \mbox{ECHo-1k} phase: gold, silver and aluminium.

Gold is the typical absorber material for MMC detectors and therefore the natural choice and the easiest one from the fabrication point of view. In pure gold, the cubic symmetry of the electric field leads to degenerate nuclear energy levels. However, when gold nuclei are in the vicinity of a holmium ion, the charge distribution is changed and the lattice is deformed, causing an electric field gradient which can induce an hyperfine splitting, therefore leading to an additional contribution to the heat capacity \cite{Herr2000}.

On the other hand, silver exhibits a very similar electron shell configuration with respect to gold, but no nuclear quadrupole moment is present and therefore no hyperfine splitting is expected due to the implanted holmium ions. Thus, silver has been investigated as host material candidate.
Additionally, the heat capacity contribution from the $^{163}$Ho ions implanted in silver is lower than for gold in the MMC operational range below $20 \, \mathrm{mK}$ \cite{Ho_Au_HC}.

Finally, aluminium allows for a larger penetration depth during the implantation process. According to simulations, the estimated mean longitudinal implantation depth is $19.5 \, \mathrm{nm}$, while in the case of gold and silver this value is $5.7 \, \mathrm{nm}$ and $7.7 \, \mathrm{nm}$, respectively \cite{implantation_sim}. Therefore it should be possible to reach higher $^{163}$Ho activities by limiting the sputtering of previously implanted $^{163}$Ho ions. In order to test this effect, an ECHo-1k chip has been implanted with aluminium as host material. A potential drawback of aluminium is the superconducting nature of this material, which could potentially compromise the detector response.

\subsection{Injection of the persistent current} 
\label{SUBSEC:MMC_operation}

\begin{figure}[h!] 
	\centering
	\includegraphics[width=0.85\textwidth]{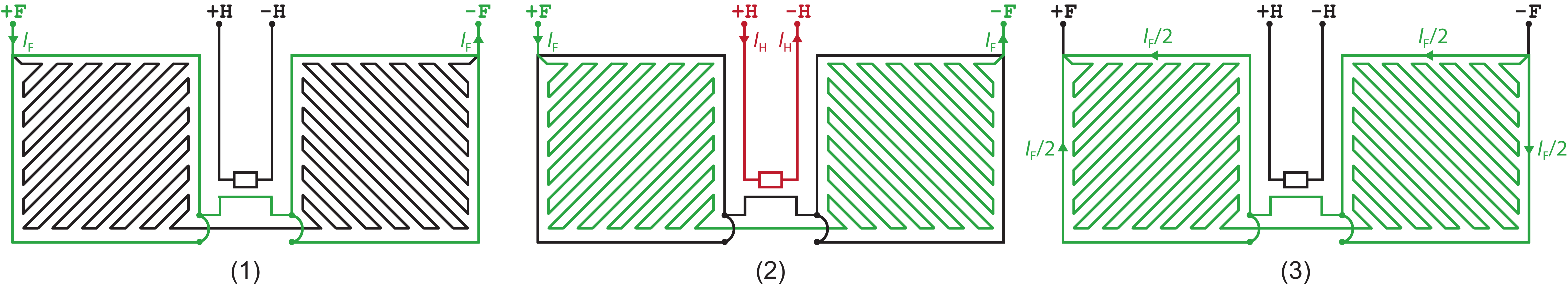}
	\caption{Schematic layout showing the procedure to inject persistent super-current into the meander-shaped pick-up coils in order to polarise the sensor. The heater element is connected to the $\pm$H lines. The meandering structures and a shorter path external to the meanders are connected in parallel to the $\pm$F lines. The steps for the current-injection are explained in detail in the text. } 
	\label{FIG:freezing_currents}
\end{figure}

The static magnetic field used to polarise the sensor is created by a persistent current circulating in the closed loop formed by the superconducting meander-shaped pick-up coils. The current can be injected using a persistent current switch, exploiting a normal-conducting element that acts as a heater which drives a small part of the closed superconducting circuit to the normal-conducting state by local Joule heating. The procedure to prepare the supercurrent is schematically shown in figure \ref{FIG:freezing_currents}:

\begin{enumerate}
    \item The field generating current ($I_\mathrm{F} \sim 30 \, \mathrm{mA}$) is injected from $+$F to $-$F. Since the flux in a superconducting loop is conserved, the major part of the current flows in the arm having lower inductance;
    
    \item The heater is activated driving a current ($I_\mathrm{H} \sim 4.0 \, \mathrm{mA}$) through the Au:Pd resistor film, characterised by a resistance of about $\mathrm{10 \, \Omega}$, as measured at millikelvin temperature. In this way, part of the niobium line is driven to the normal conducting regime, showing a finite resistance. Consequently, the field generating current changes its path and will then flow through the large inductance formed by the meander-shaped pick-up coils, which are still in the superconducting state;
    
    \item When the heater is deactivated, all the lines turn superconducting again and due to the flux conservation in superconducting loops, the field generating current keeps flowing in the meander-shaped pick-up coils.
    
\end{enumerate}

The functioning of the persistent current switch is fundamental for the operation of the detectors and therefore the switching point of the heater elements has been tested and carefully characterised.

All the pick-up coils belonging to one quarter of the chip are connected in series. The same is valid for the heater elements of one quarter. Therefore, the persistent current is injected into all the detectors belonging to one quarter (i.e.~9 detector channels) at a time. The pads connecting the current sources for the heater and the field lines are located in the right periphery of the chip (as shown in figure \ref{SUBFIG:ECHo-1k_chip}).

\subsection{Detector thermodynamic properties}

The detector heat capacity $C$ has a direct impact on the detector performances, in particular on signal amplitude $A$ and on the fundamental limit of the energy resolution $\Delta E_{\mathrm{FWHM}}$, according to the relations:

\begin{equation} \label{EQ:amplitude}
 A \propto \Delta T = E/C
\end{equation} 

\begin{equation} \label{EQ:energy_res}
 \Delta E_{\mathrm{FWHM}} \propto \sqrt{4 k_\mathrm{B} T^2 C}
\end{equation}

\noindent where $\Delta T$ is the temperature change that follows the particle interaction, $k_\mathrm{B}$ is the Boltzmann constant and $T$ is the operational temperature.

The detector heat capacity contains three contributions: the heat capacity of the sensor $C_{\mathrm{s}}$, the heat capacity of the absorbers $C_{\mathrm{a}}$ and the heat capacity of the $^{163}$Ho atoms $C_{\mathrm{Ho}}$ \cite{Ho_Au_HC}. 
The heat capacity of the absorber layers is calculated from the absorber geometry, while the heat capacity of the Ag:Er sensor with given geometry and erbium concentration can be derived from dedicated sensor simulations based on the diagonalisation of the Hamiltonian of randomly distributed spins in the metal lattice \cite{Enss2000}. Finally, the heat capacity of the implanted $^{163}$Ho ions can be estimated knowing the specific heat of holmium in the host material \cite{Ho_Au_HC} and the number of implanted ions. \\
The design values for the ECHo-1k detector, considering silver as host material, a Ag:Er concentration of $445 \, \mathrm{ppm}$ for the sensor fabrication and a temperature $T = 20 \, \mathrm{mK}$ are $C_{\mathrm{a}} = 0.46 \, \mathrm{pJ/K}$, $C_{\mathrm{s}} = 3.22 \, \mathrm{pJ/K}$ and $C_{\mathrm{Ho}} = 0.02 \, \mathrm{pJ/K}$, assuming an activity of $1 \, \mathrm{Bq}$.

The detector heat capacity together with the thermal conductance $G$ of the thermal link that connects the sensor to the thermalisation bath determine the decay time $ \tau_\mathrm{d}$ of the temperature signal:

\begin{equation} \label{EQ:decay_time}
    \tau_\mathrm{d} \approx \frac{C}{G}
\end{equation}

For the ECHo-1k detectors, the thermal link with the smallest thermal conductance consists of the gold stripe with dimensions $\mathrm{48.5 \, \upmu m \times 5 \, \upmu m \times 300 \, nm}$ (figure \ref{FIG:thermal_link}), which corresponds to $G = 1.08 \times 10^{-9} \, \mathrm{J/K/s}$ at $20 \, \mathrm{mK}$.

\section{Characterisation of the ECHo-1k detector array} \label{SEC:characterisation}

After fabrication, a number of ECHo-1k detector chips have been tested and fully characterised, in order to proof their functionality, verify the design values and evaluate operational parameters. The resulting positive outcomes allowed to proceed with the $^{163}$Ho ion-implantation. 

\subsection{Characterisation at room temperature}

Tests at room temperature are relatively fast and easy to perform, therefore they can be a powerful tool to verify the quality of the single detector on a chip or wafer scale.

\begin{figure}[h!] 
    \centering
    \begin{subfigure}[b]{0.4\textwidth}
        \centering
        \includegraphics[trim=12 0 12 0,clip,width=\linewidth]{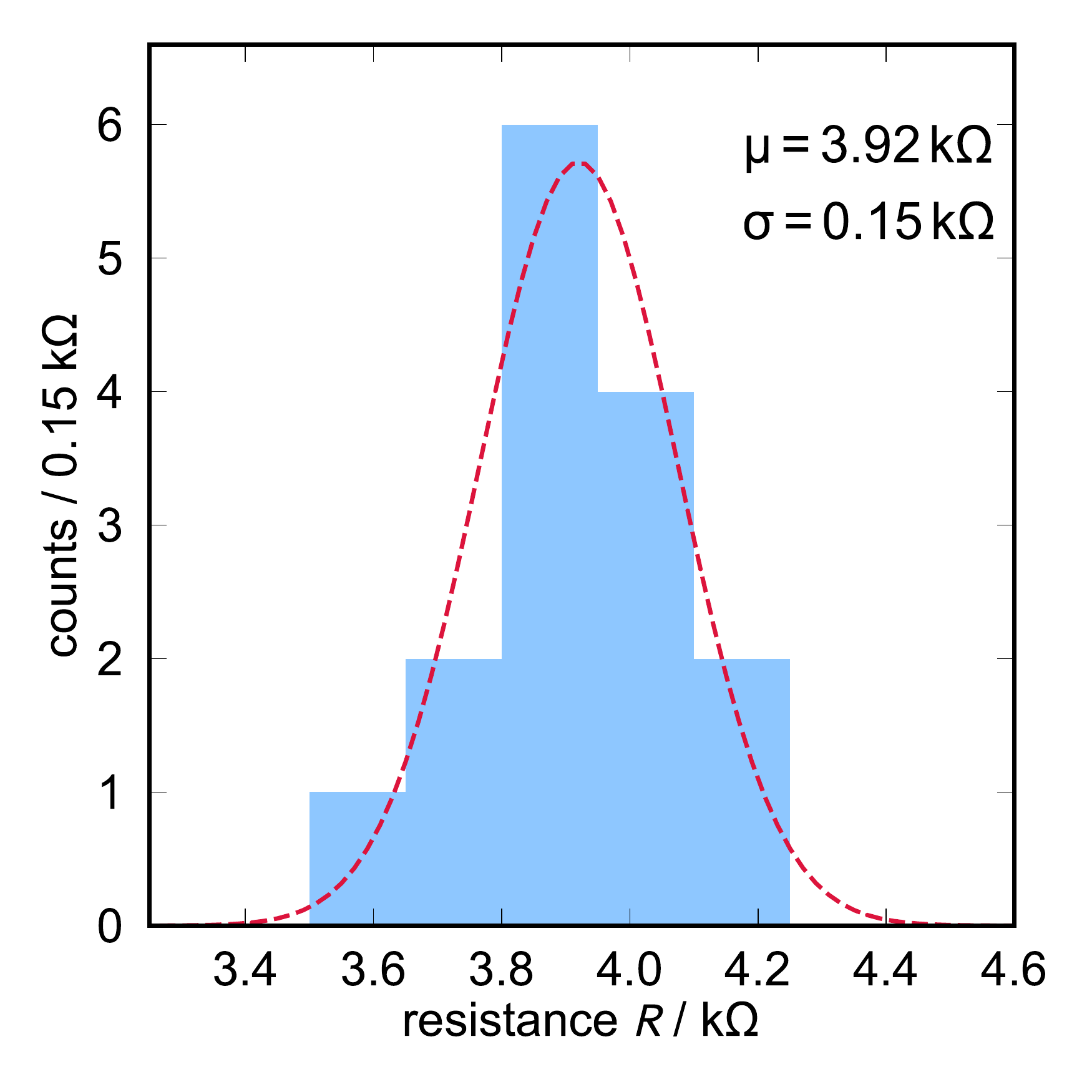}
        \caption{} \label{SUBFIG:Rf_300K}
    \end{subfigure}
    \hfill
    \begin{subfigure}[b]{0.4\textwidth} 
        \centering
        \includegraphics[trim=12 0 12 0,clip,width=\linewidth]{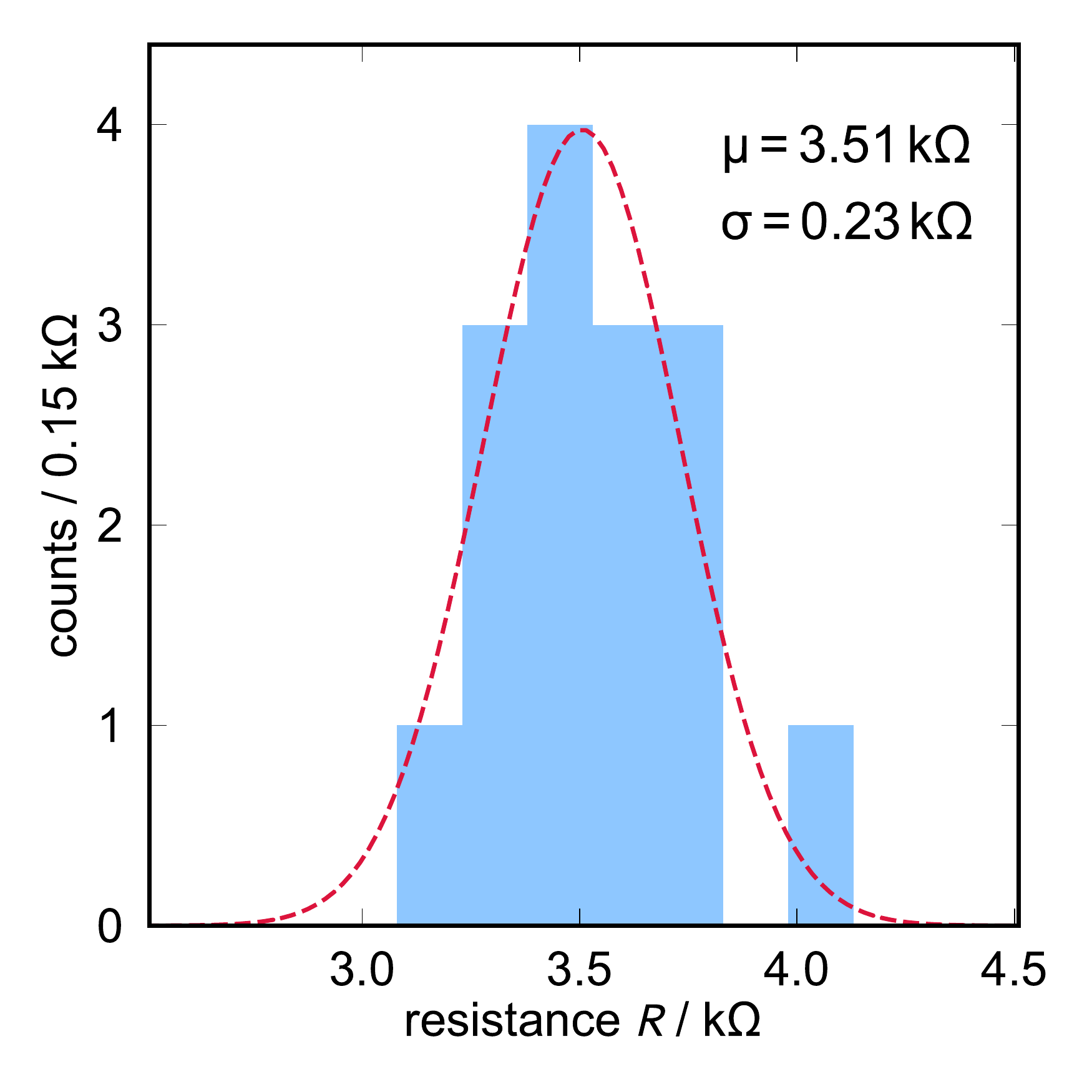}
        \caption{} \label{SUBFIG:Rh_300K}
    \end{subfigure}

    \caption{Histograms of the resistances at room temperature of the $\pm$F lines \textbf{(a)} and $\pm$H lines \textbf{(b)} of one quadrant of 15 ECHo-1k chips. The Gaussian fit is shown in red and the corresponding mean $\upmu$ and standard deviation $\upsigma$ are reported.}

    \label{FIG:R_300K}
\end{figure}

As first step the chips undergo an optical inspection to exclude evident fabrication issues or damages that can compromise the functionality of the chip. Among others, the continuity of the niobium lines, the stability of the absorbers and the condition of the heater elements and of the bond-pads are important aspects to be verified.
After that, the resistances of the niobium lines can be measured at room tem\-pe\-ra\-ture contacting the corresponding bond-pads with a needle-probe tool, checking for possible shorts.
The resistances of the $\pm$F lines connected to the pick-up coils and the resistances of the $\pm$H lines connected to the heater elements have been measured for 15 ECHo-1k chips belonging to the same wafer and the resulting values are shown in the histograms of figure \ref{FIG:R_300K}. The scatter of the resistance values can be attributed to a variation up to 10 \% in the thickness of the deposited niobium layer over the wafer.

Resistance values in the range of $\mathrm{3 - 5 \, k\Omega}$ at room temperature prove the continuity of the corresponding conducting structures. 
The continuity of the niobium lines that connect the detector pick-up coils to the SQUID pads, to which the SQUID chip will be wire-bonded, is also tested at room temperature and the resistance between all the pads is checked to exclude the presence of shorts on the chip.

\subsection{Characterisation at 4 K}
\label{SUBSEC:ECHo-1k_characterisation_4K}

Several properties of the ECHo-1k detector can be investigated in a temperature range around $4 \, \mathrm{K}$.
Two experimental approaches can be pursued to reach this temperature, namely immersing the set-up in a liquid helium bath or exploiting the first cooling step of a dry dilution refrigerator based on a two-stage pulse tube cryocooler.
Both these methods have been employed for the characterisation measurements that are presented in the following.

\subsubsection*{Persistent current switch}

\begin{figure}[h!] 
    \centering
    \begin{subfigure}[b]{0.3\textwidth}
        \centering
        \includegraphics[height=.31\textheight]{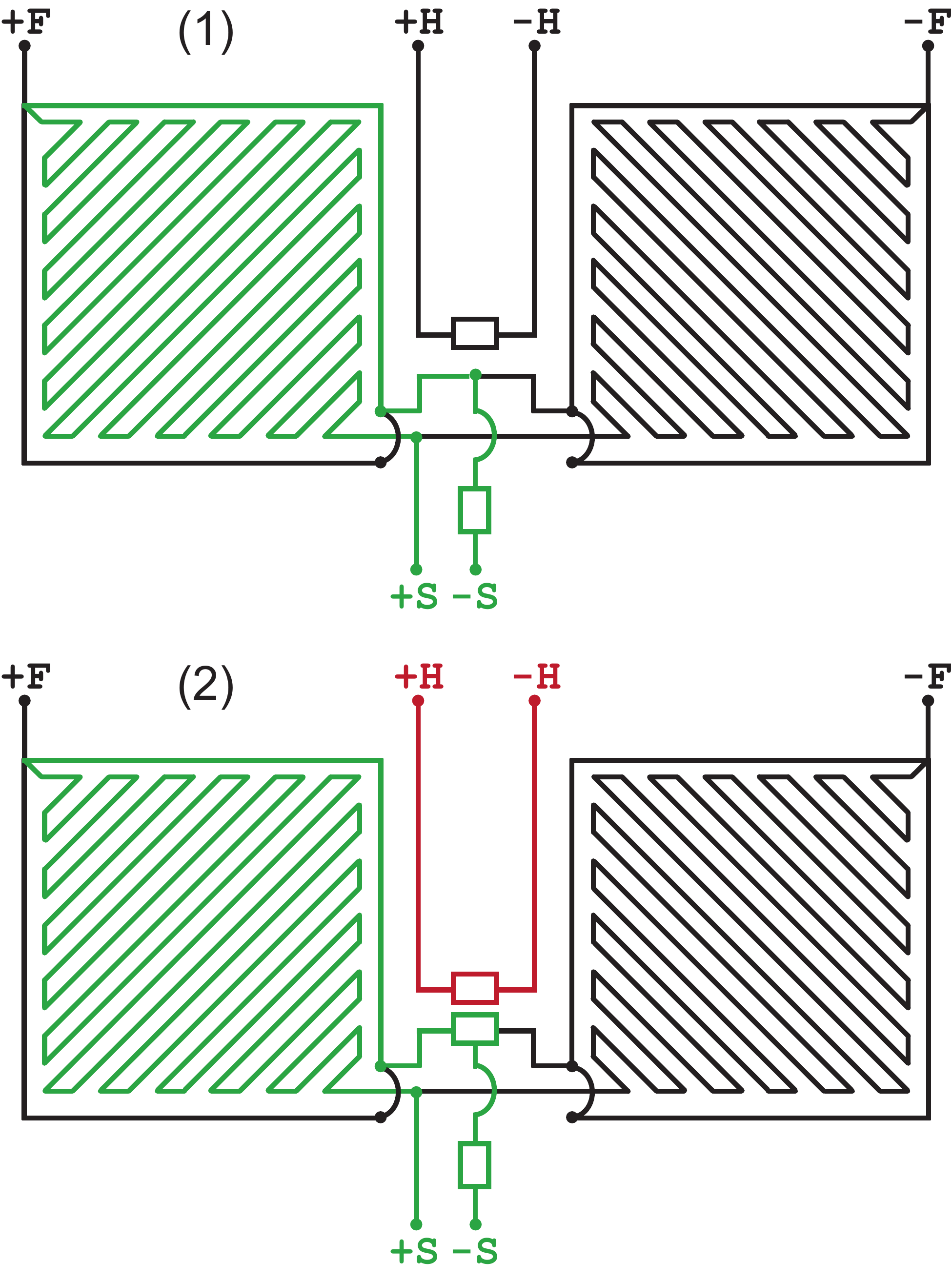}
       \caption{} \label{SUBFIG:switch_point_meas}
    \end{subfigure}
    \hfill
    \begin{subfigure}[b]{0.4\textwidth} 
        \centering
        \includegraphics[trim=12 0 12 0,clip,width=\linewidth]{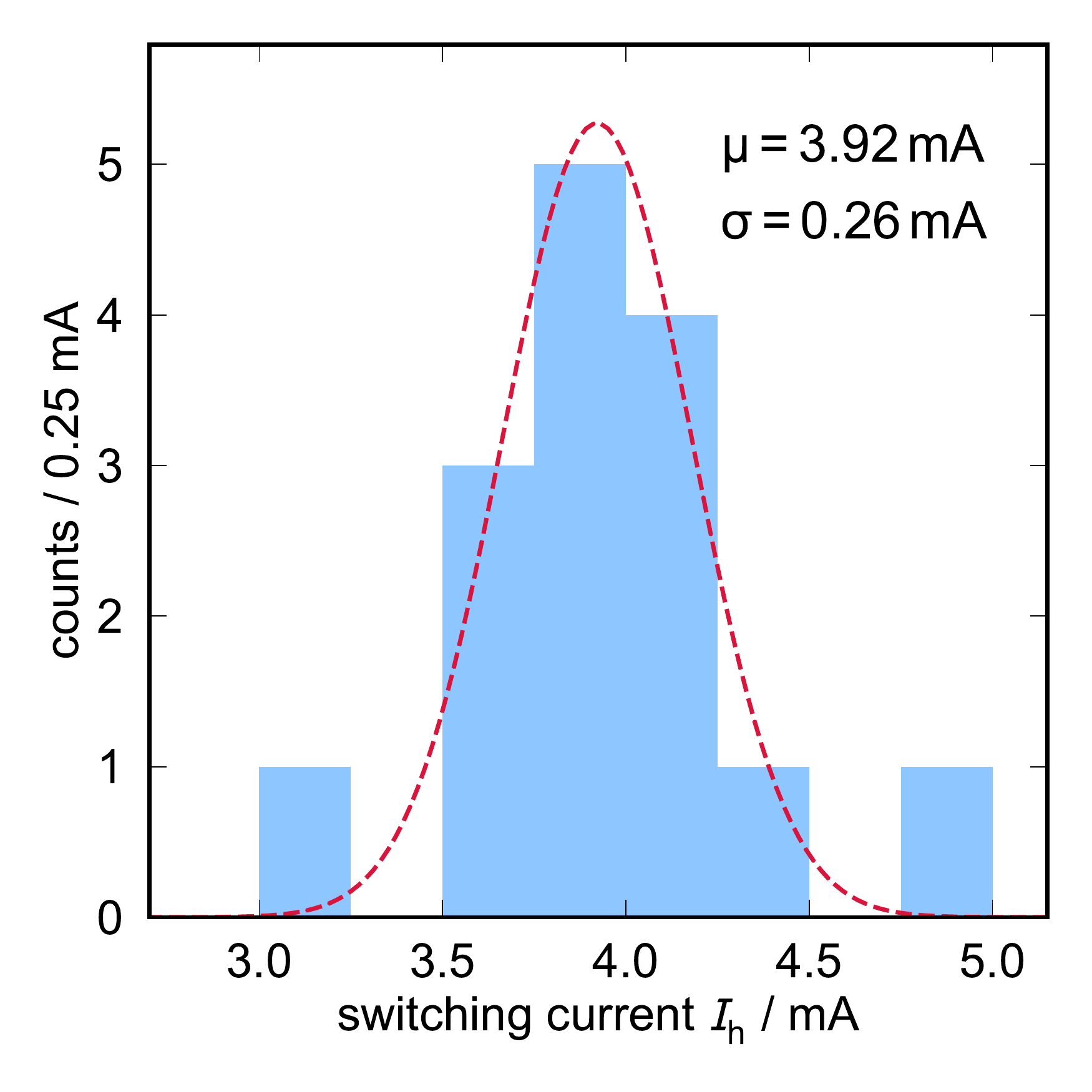}
        \caption{} \label{SUBFIG:hist_I_switch}
    \end{subfigure}

    \caption{\textbf{a)} Layout of the measurement method to determine the current to activate the heater switch. \textbf{b)} Histogram of the heater switch current values measured for one quadrant of 15 ECHo-1k chip; the distribution is fitted with a Gaussian function and the resulting mean $\upmu$ and standard deviation $\upsigma$ are reported.}

    \label{FIG:switch_point}
\end{figure}

In order to probe the chip functionality, it is crucial to check the performance of the persistent current switch and to determine the current value that activates it. The experimental method is based on the measurement of the resistance of the lines connected to the SQUID read-out pads at a temperature of $\mathrm{4.2 \, K}$. As depicted in figure \ref{SUBFIG:switch_point_meas}, the SQUID read-out pads $\pm$S are connected to the pick-up coils, therefore forming a niobium superconducting circuit. 
Increasing currents are injected through the heater contacts monitoring the resistance between the pads $\pm$S. When the current through the heater generates enough power to locally break the superconductivity, part of the probed circuit enters the normal-conducting regime, creating a non-zero resistance that can be measured. The minimal current leading to the resistance jump is taken as heater switching current.

The heater switching current values have been determined for 15 ECHo-1k chips and are reported in a histogram in figure \ref{SUBFIG:hist_I_switch}. The average heater switching current value is $I_\mathrm{h,mean} = \mathrm{3.92 \, mA}$ in a liquid helium bath.

\subsubsection*{Inductance of the pick-up coil}

\begin{figure}[h!] 
	\centering
	\includegraphics[width=.45\textwidth]{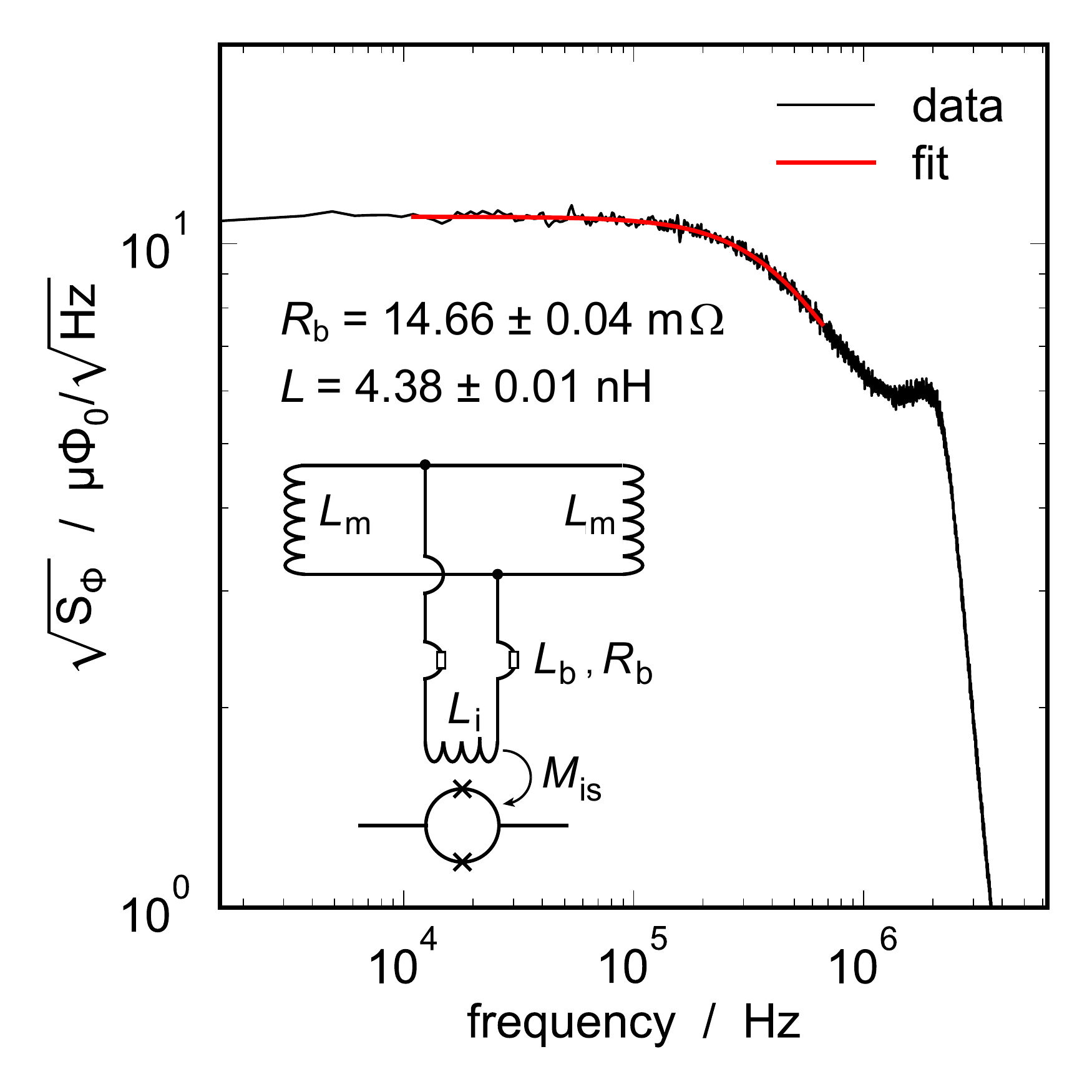}
	\caption{Measured spectral flux density in the SQUID for an ECHo-1k detector channel. The data is shown in black while the fit with the parameters $R_\mathrm{b}$ and $L$ quoted in the figure is shown in red. The inset shows the layout of the corresponding circuit. } 
	\label{FIG:noise}
\end{figure}

The inductance of the niobium superconducting meander-shaped pick-up coils is an essential parameter to calculate the flux coupling between detector and SQUID \cite{Fle2005}.
This value can be determined from a SQUID flux noise measurement performed at $T \approx \mathrm{4 \, K}$. At this temperature the aluminium bon\-ding wires connecting the pick-up coil and the input coil of the SQUID are normal-conducting, while the niobium meander-shaped pick-up coils as well as the input coil of the SQUID are in the superconducting regime. Therefore, the resistance of the bonding wires $R_\mathrm{b}$ together with the inductance of the circuit $L$ constitute a low-pass filter with a cut-off frequency $f_\mathrm{c} = R_\mathrm{b}/(2 \pi L) $. The corresponding circuit is shown in the inset of figure \ref{FIG:noise}. 

The expected total noise spectral density is given by:

\begin{equation} \label{EQ:noise}
    S_{\mathrm{\Phi_\mathrm{S}}} = M_{\mathrm{is}}^2 \frac{4 k_\mathrm{B} T}{R_\mathrm{b}} \frac{1}{1+(f/f_\mathrm{c})^2} + S_{\mathrm{\Phi_\mathrm{S}}}^\mathrm{S,w}
\end{equation}

\noindent where $M_\mathrm{is}$ is the mutual inductance between the pick-up coil and the SQUID and $S_\mathrm{\Phi_\mathrm{S}}^\mathrm{S,w}$ is the white noise of the SQUID.
The plateau at low frequency is given by the Johnson noise of the bonding wires.

The spectral flux density measured in the SQUID, with a detector channel connected to it, is shown in figure \ref{FIG:noise} together with a fit performed using equation \ref{EQ:noise}, where $R_\mathrm{b}$ and $L$ are free parameters. From the value of the total inductance $L = \mathrm{4.38 \pm 0.01 \, nH}$ given by the fit it is possible to estimate the inductance of a single meander-shaped pick-up coil $L_\mathrm{m}$:

\begin{equation} \label{EQ:L_m}
    L_\mathrm{m} = 2 \cdot (L-L_\mathrm{i}-L_\mathrm{b})
\end{equation}

\noindent where $L_\mathrm{i}$ is the inductance of the SQUID input coil, which by design is $1.8 \, \mathrm{nH}$, and $L_\mathrm{b}$ is the inductance of the bonding wires, which can be estimated from the corresponding resistance assuming the relation $L_\mathrm{b} \approx c \cdot R_\mathrm{b} $, with $c = 0.10 \pm 0.02 \, \mathrm{nH/m \Omega}$ and $R_\mathrm{b} = 14.66 \pm 0.04 \, \mathrm{m\Omega}$ from the fit. The value of $c$ is based on the outcome of noise measurements performed varying the length of the aluminium wire-bonds and its uncertainty is the dominant one in the estimation of the inductance of the meander-shaped pick-up coil.
This calculation leads to a pick-up coil inductance value of $L_\mathrm{m} = 2.2 \pm 0.6 \, \mathrm{nH}$.
The design value\footnote{The pick-up coil inductance has been simulated with InductEx, \url{http://www0.sun.ac.za/ix}.} of the pick-up coil inductance is $L_\mathrm{m,d} = \mathrm{2.27 \, nH}$, which is consistent with the experimentally determined value.

The characterisation measurements performed at room temperature and at $4 \, \mathrm{K}$ have shown that the resistance values measured at room temperature are valid parameters to verify the detector functionality at millikelvin. Several ECHo-1k detector chips have been tested at room temperature and the ones with the expected resistance values have been also operated at millikelvin with a good performance.
Furthermore, it has been demonstrated that resistance values of the $\pm$H lines measured at $\mathrm{4.2 \, K}$ that lay within three standard deviations from the average value lead to a proper behaviour of the persistent current switch at millikelvin tem\-pe\-ra\-ture.

\subsection{Characterisation at millikelvin temperature}
\label{SUBSEC:ECHo-1k_characterisation_mK}

In order to determine the properties of the ECHo-1k detector array at low temperature, the cryogenic platform dedicated to the ECHo experiment and the parallel read-out chain described in \cite{ECHo_readout} have been used. The detectors have been fully characterised in terms of pulse shape, magnetisation response, energy resolution and $^{163}$Ho activity.

\subsubsection*{Magnetisation response}

The detector response of MMCs is based on the change of magnetisation in the sensor depending on the detector temperature, as described in section \ref{SEC:MMCs}. \\
The magnetisation response is measured stepwise changing the tem\-pe\-ra\-ture of the mixing chamber plate (MXC) and acquiring the SQUID voltage output of a non-gradiometric channel. For each temperature step a waiting time of at least 30 minutes is set before starting the SQUID voltage measurement, to allow the full set-up connected to the mixing chamber plate to have a stable temperature value. In this way, the temperature of the detector can be reliably estimated from the MXC temperature. Since the voltage range of the ADC is limited, an automatic reset\footnote{When the SQUID output voltage exceeds the threshold value set by the user, the SQUID is set in open loop and then locked back to FLL mode.} of the SQUID is applied to guarantee that the SQUID voltage output does not exit the ADC range. 
The output of the MXC thermometer and the voltage output of the SQUID are synchronised and recorded continuously. 
The flux in the SQUID $\Phi_\mathrm{S}$ is proportional to the SQUID output voltage, according to the relation:

\begin{equation}
    \Phi_\mathrm{S} = V \cdot \frac{M_\mathrm{fb}}{R_\mathrm{fb}} \, .
\end{equation}

\noindent where $M_\mathrm{fb}$ is the mutual inductance between the feedback coil of the two-stage SQUID set-up and the front-end SQUID and $R_\mathrm{fb}$ is the feedback re\-sis\-tance of the feedback loop.

\begin{figure}[h!] 
	\centering
	\includegraphics[width=.65\textwidth]{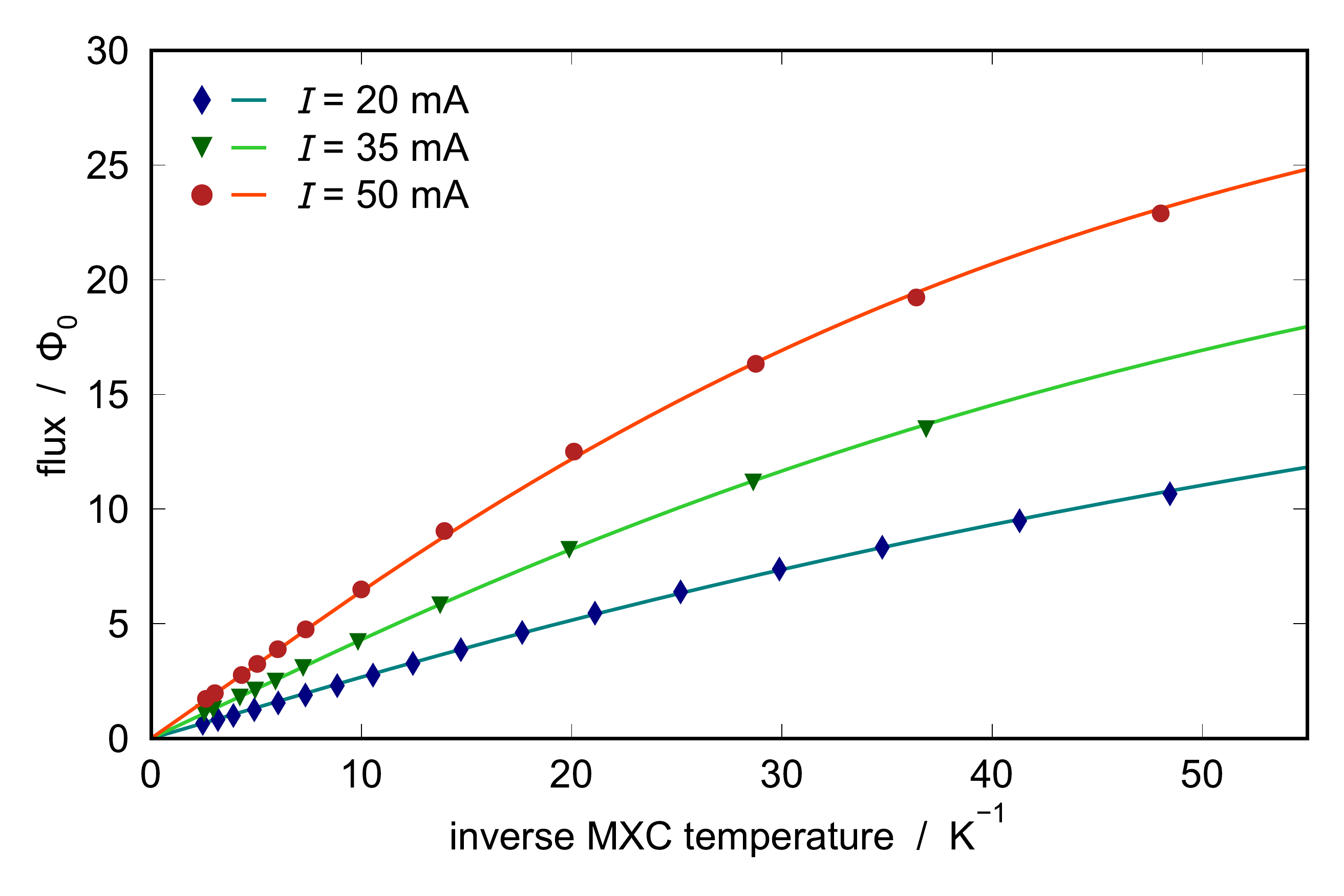}
	\caption{Magnetisation response of the ECHo-1k detector for three different persistent current values ($20 \, \mathrm{mA}$, $35 \, \mathrm{mA}$ and $50 \, \mathrm{mA}$) in the detector pick-up coils. The experimental data points are shown with markers and the theoretical expectations are shown with solid lines.} 
	\label{FIG:magnetisation_ECHo-1k}
\end{figure}

The resulting data points for three different values of the persistent current in the pick-up coils, namely $20 \, \mathrm{mA}$, $35 \, \mathrm{mA}$ and $50 \, \mathrm{mA}$, are shown in figure \ref{FIG:magnetisation_ECHo-1k} together with the corresponding expectations\footnote{For ECHo-10k sensor simulations a concentration of $440 \, \mathrm{ppm}$, as experimentally determined, and a $\alpha$ parameter of 13.5 are used.}. The sensor thickness has been estimated to be about $1 \, \mathrm{\upmu m}$ from the measured sensor heat capacity and this value has been confirmed by a direct thickness measurement\footnote{The thickness measurement was performed with Dektak XT}. 
The latter can be extracted from the amplitude of the signals generated by the $5.9 \, \mathrm{keV}$ photons from an external $^{55}$Fe calibration source, which are recorded with the detector pixels that are equipped only with the sensor and no absorber. The resulting sensor thickness value has been used as input parameter for the numerical simulations. The simulated magnetisation curves describe the experimental curves well, as shown in figure \ref{FIG:magnetisation_ECHo-1k}. \\

\subsubsection*{Pulse shape}

\begin{figure}[h!] 
    \centering
    \begin{subfigure}[b]{0.7\textwidth}
        \centering
        \includegraphics[width=1\linewidth]{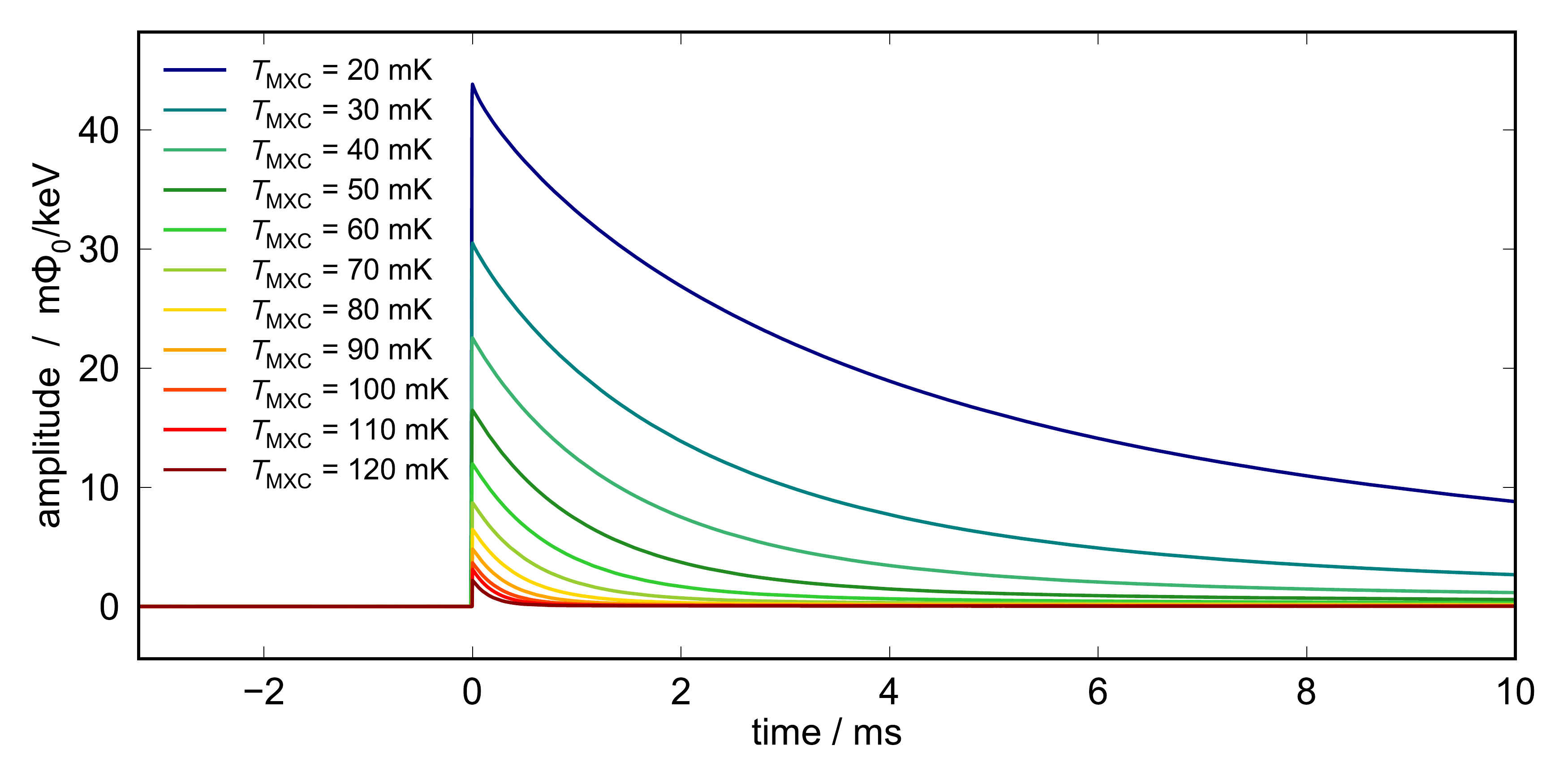}
        \caption{} \label{SUBFIG:temp_scan_lin}
    \end{subfigure}
    \hfill
    \begin{subfigure}[b]{0.7\textwidth} 
        \centering
        \includegraphics[width=1\linewidth]{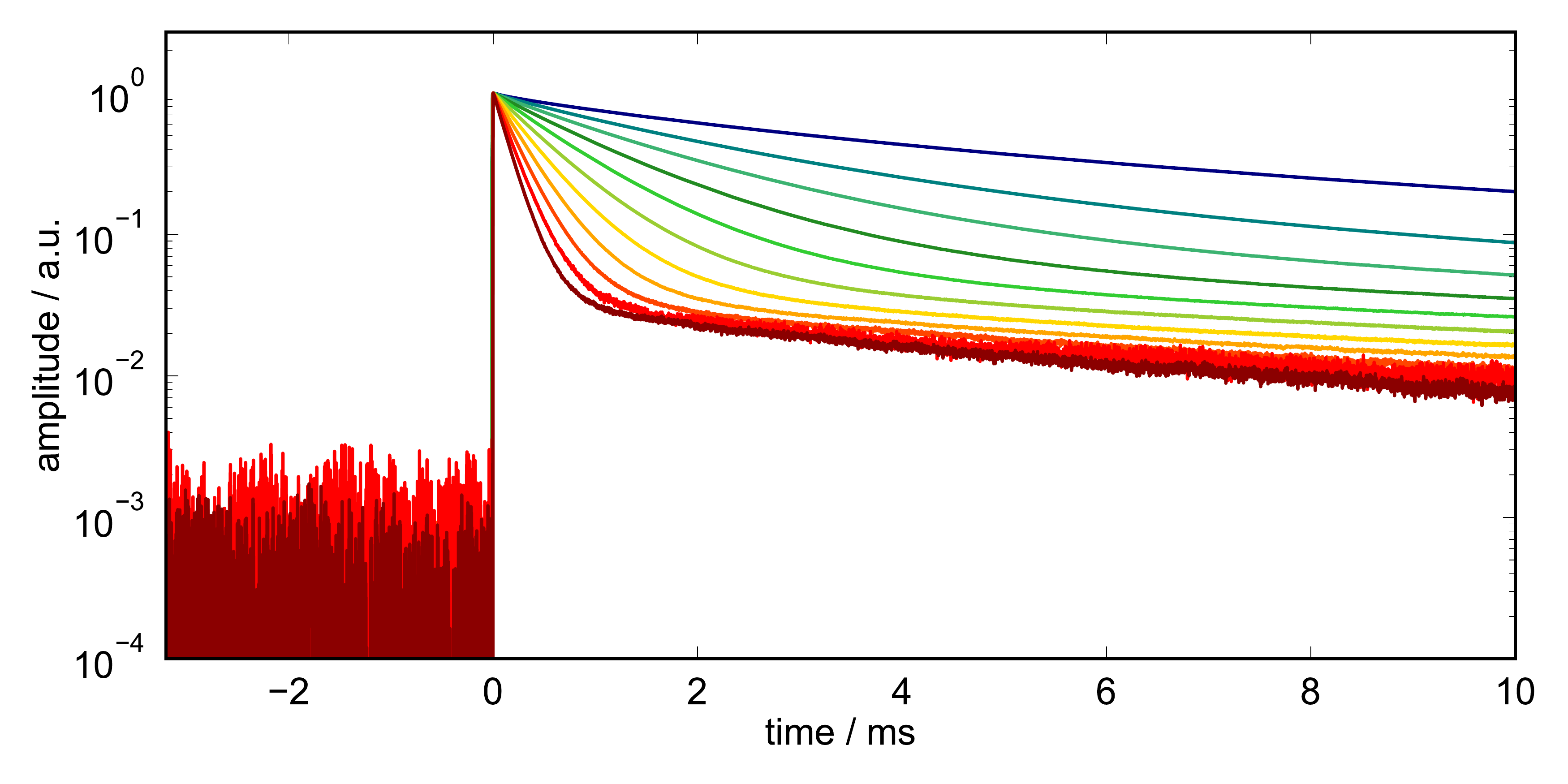}
        \caption{} \label{SUBFIG:temp_scan_log}
    \end{subfigure}

    \caption{Signals corresponding to the $^{55}$Fe K$_\alpha$ line recorded with a non-implanted ECHo-1k detector with two absorber layers having a current of $35 \, \mathrm{mA}$ in the superconducting pick-up coils. In \textbf{(a)} the vertical axis is shown in linear scale. In \textbf{(b)} the pulses are scaled to the same amplitude and the vertical axis is shown in logarithmic scale. The legend in \textbf{(a)} reports the MXC temperature $T_{\mathrm{MXC}}$ and is valid also for \textbf{(b)}.}

    \label{FIG:temp_scan}
\end{figure}

As a first step after fabrication, the ECHo-1k detectors with only one absorber layer have been tested and shortly characterised with an external sealed $^{55}$Fe source. 
The positive outcomes of this measurements allowed to proceed with the implantation process, enclosing $^{163}$Ho in the first absorber layer and depositing the second absorber layer on top. \\
Two ECHo-1k detector chips have been implanted according to the implantation procedure described in section \ref{SEC:detector_design}. 
A study of the temperature dependent pulse shape has been performed by acquiring the detector signals at different temperatures of the mixing chamber plate of the cryostat.
The signal shape has been measured with different currents in the superconducting meander-shaped pick-up coils, namely $20 \, \mathrm{mA}$, $35 \, \mathrm{mA}$ and $50 \, \mathrm{mA}$. 
Figure \ref{FIG:temp_scan} shows the signals corresponding to the $^{55}$Fe K$_\alpha$ photons stopped in a non-implanted pixel with both the absorber layers. The signals are acquired at different temperatures with a persistent current of $35 \, \mathrm{mA}$ in the pick-up coils.

As expected, the signal amplitudes are decreasing at higher temperatures due to the increase of the detector heat capacity. The decay time becomes faster at higher temperatures, as visible in figure \ref{SUBFIG:temp_scan_log}. This is caused by the increase of the thermal conductance of the thermal link that connects the detector to the thermal reservoir. 

\begin{figure}[h!] 
	\centering
	\includegraphics[width=.45\textwidth]{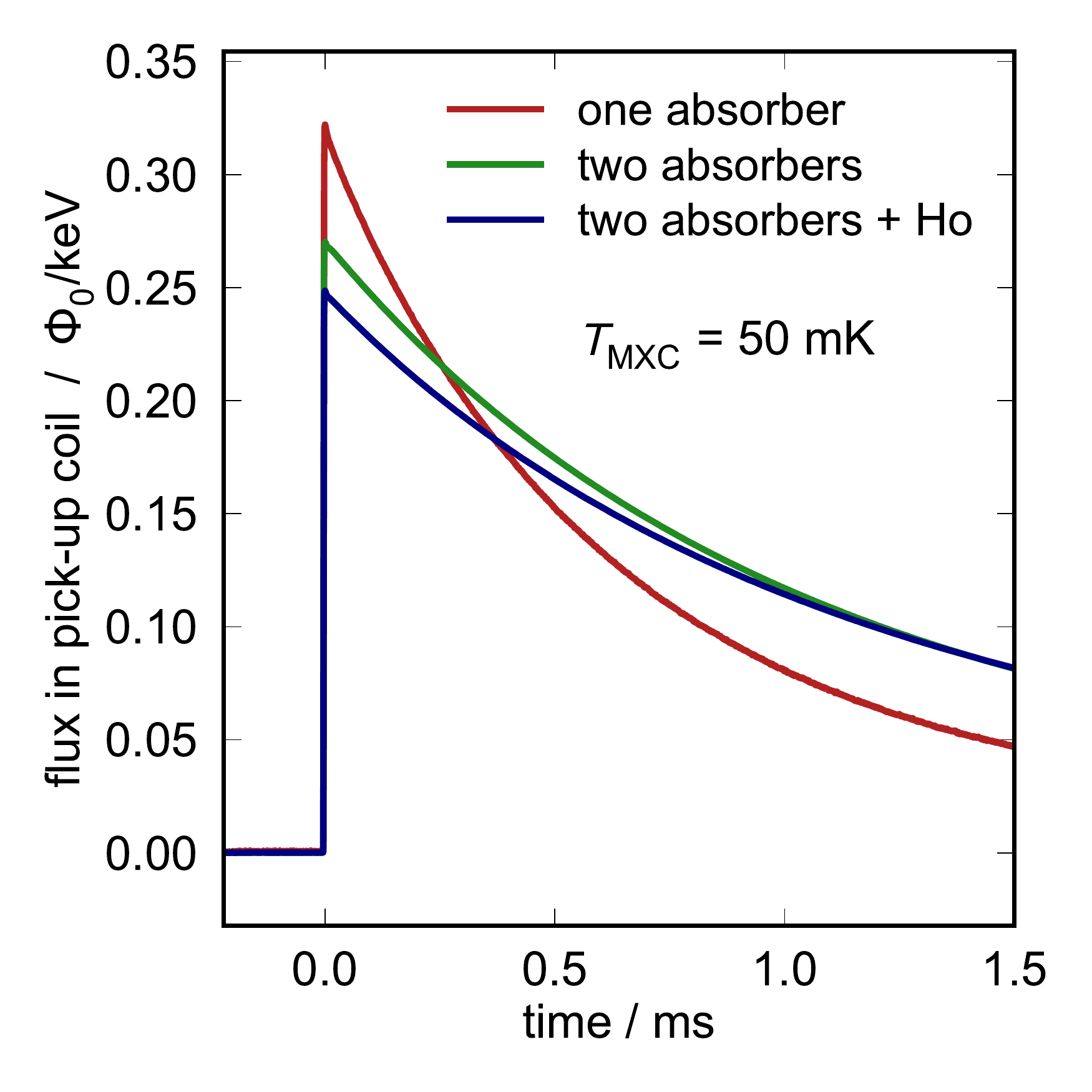}
	\caption{Comparison between the detector responses in terms of flux in the pick-up coil after fabrication (with one absorber), after implantation (with two absorbers) with $^{163}$Ho and without $^{163}$Ho at a MXC temperature of $50 \, \mathrm{mK}$. The persistent current in the pick-up coil is $20 \, \mathrm{mA}$.} 
	\label{FIG:comparison_before_after_impl}
\end{figure}

\newpage
\paragraph{Effects of implanted holmium}
Comparing the detector response before and after the $^{163}$Ho implantation, changes in amplitude and in decay time are observed. In figure \ref{FIG:comparison_before_after_impl} the detector responses in terms of flux change in the pick-up coil for the following detector pixels are compared:
 
\begin{itemize}
    \item a detector pixel before implantation, i.e.~with only one absorber layer and with no $^{163}$Ho;
    \item a non-implanted detector pixel, i.e.~with two absorber layers but no $^{163}$Ho enclosed in the absorbers;
    \item a detector pixel after implantation with implanted source, i.e.~with two absorber layers and $^{163}$Ho enclosed in the absorbers.
\end{itemize}

Since additional heat capacity contributions are carried by the second absorber layer and by the $^{163}$Ho ions, the amplitude of the detector response becomes smaller, according to equation \ref{EQ:amplitude}. 
The difference in amplitude between two pixels, both with two absorber layers, with and without $^{163}$Ho can be exploited to estimate the extra heat capacity due to the implanted $^{163}$Ho ions and therefore to calculate the heat capacity per ion of $^{163}$Ho implanted in a specific host material \cite{Ho_Au_HC}.

\paragraph{Signal decay}
Experimentally, the signal decay of the ECHo-1k detector at different temperatures can be empirically described with a fit with a multi-exponential function:

\begin{equation} \label{EQ:multiexp}
   f(t) = \sum_{i=1}^{4} a_i \cdot e^{-t/\tau_i} \, .
\end{equation}

\begin{figure}[h!] 
    \centering
    \begin{subfigure}[b]{0.42\textwidth}
        \centering
        \includegraphics[trim=12 0 12 0,clip,width=\linewidth]{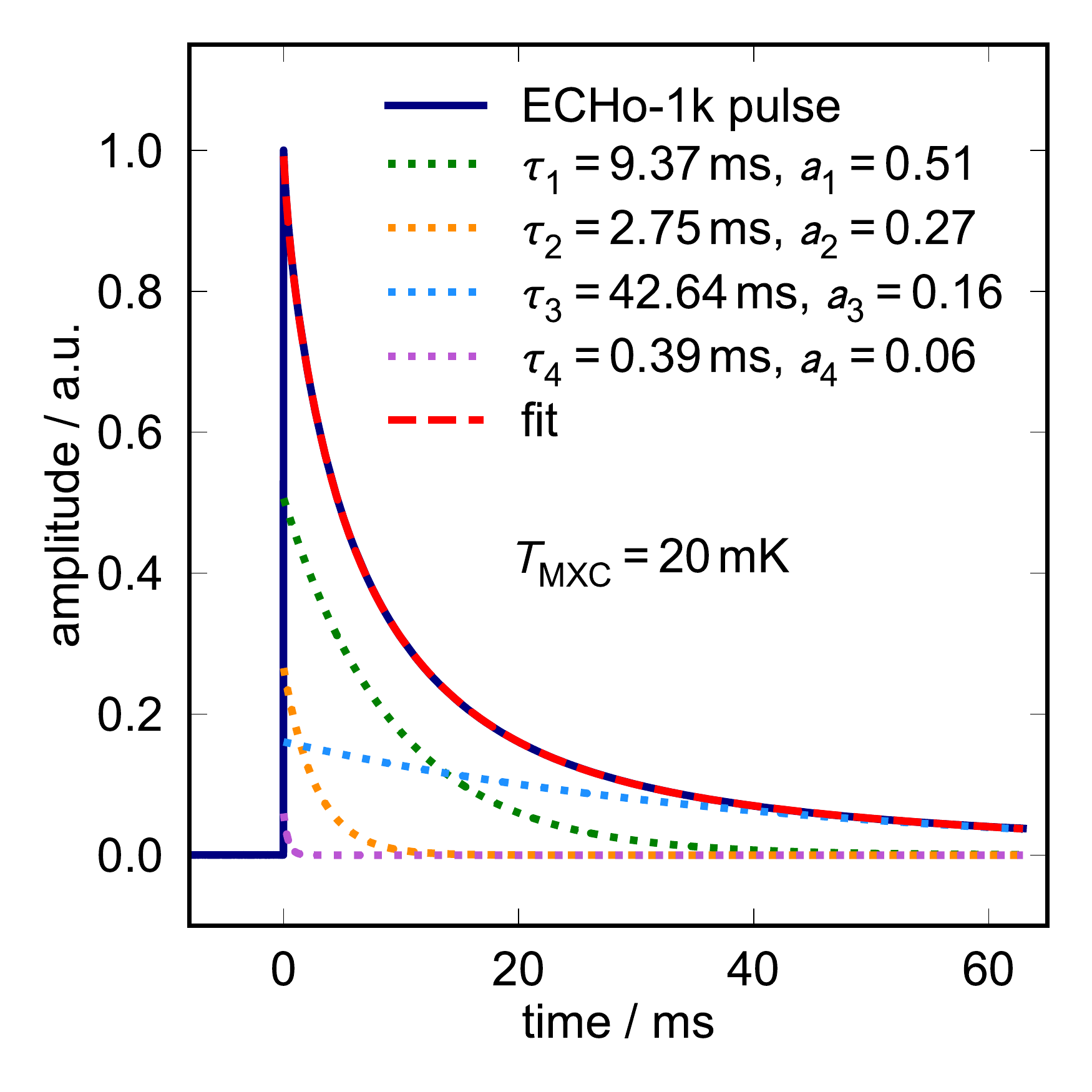}
        \caption{} \label{SUBFIG:multiexp_fit_lin}
    \end{subfigure}
    \hfill
    \begin{subfigure}[b]{0.42\textwidth}
        \centering
        \includegraphics[trim=12 0 12 0,clip,width=\linewidth]{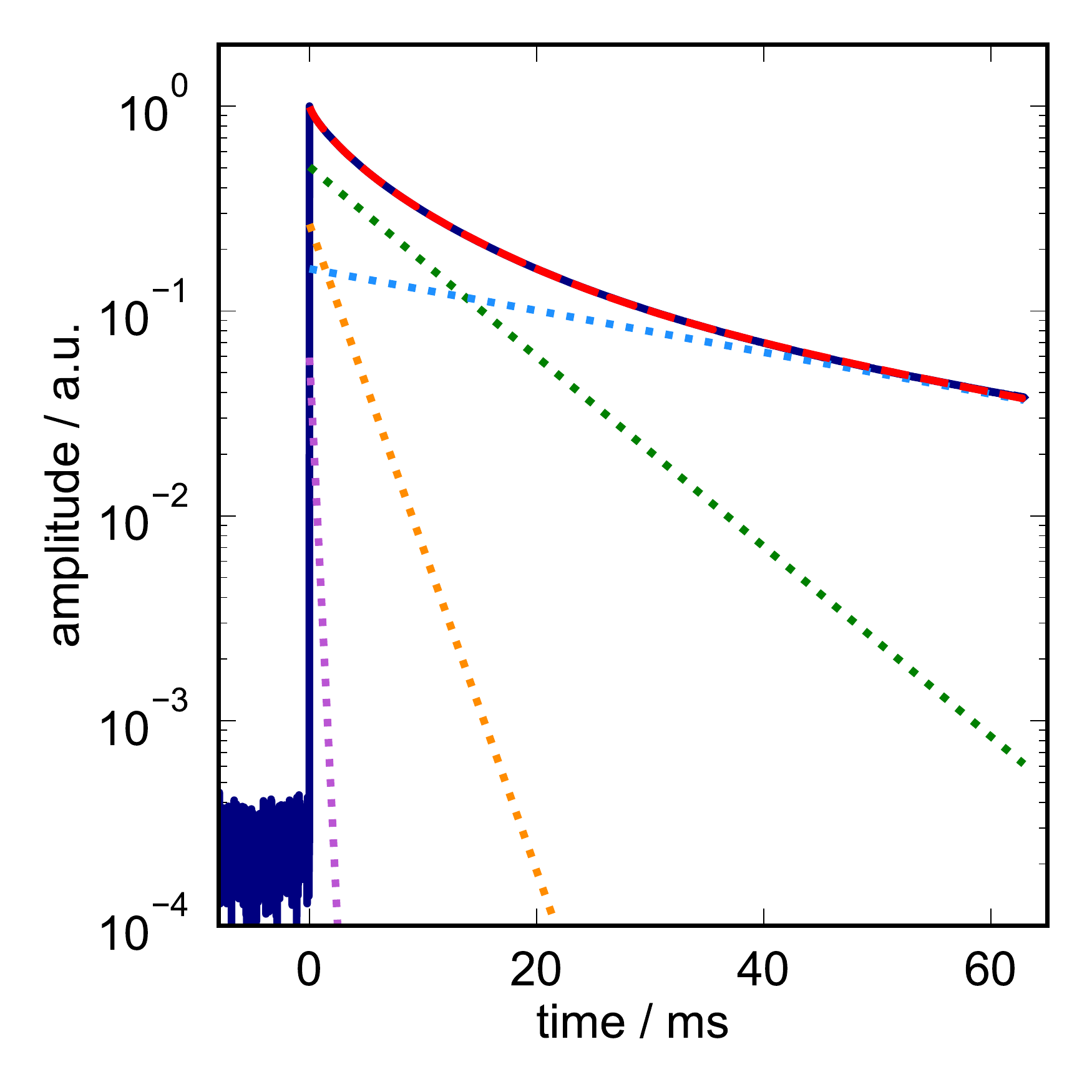}
        \caption{} \label{SUBFIG:multiexp_fit_log}
    \end{subfigure}

    \caption{Detector signal with amplitude scaled to one, acquired at $T_{MXC} = 20 \, \mathrm{mK}$ with an overlaid fit performed using equation \ref{EQ:multiexp}, in linear scale in \textbf{(a)} and in logarithmic scale in \textbf{(b)}. The four decay components are individually plotted with dashed lines. The legend of (a) is also valid for (b).}
    \label{FIG:multiexp_fit}
\end{figure}

Figure \ref{FIG:multiexp_fit} shows an example fit of a signal acquired setting the MXC temperature at $20 \, \mathrm{mK}$\footnote{The fit excludes the first $1-3 \, \upmu$s where a fast decay component probably associated to the read-out electronics is observed. Thus, the amplitudes of the exponential terms do not sum exactly to unity.}. 
According to the results of a $\chi^2$ goodness of fit test, at least four decay constants are necessary to precisely describe the decay, independently from the temperature in the range between $20 \, \mathrm{mK}$ and $140 \, \mathrm{mK}$. 
This feature points towards the hypothesis that additional thermodynamic sub-systems are present. A possible interpretation is that the complex thermalisation design of the ECHo-1k chip (figure \ref{FIG:thermal_link}) generates multiple thermalisation channels with different time constants. Furthermore, the presence of tunnelling systems in the isolation layers could contribute to generate a slow decay component. 
Long decay times do not represent an issue for the ECHo-1k experiment as the planned rate is about one event per second, which is important to limit the pile-up fraction.

\begin{figure}[h!] 
    \centering
    \begin{subfigure}[b]{0.32\textwidth}
        \centering
        \includegraphics[trim=12 0 12 0,clip,width=\linewidth]{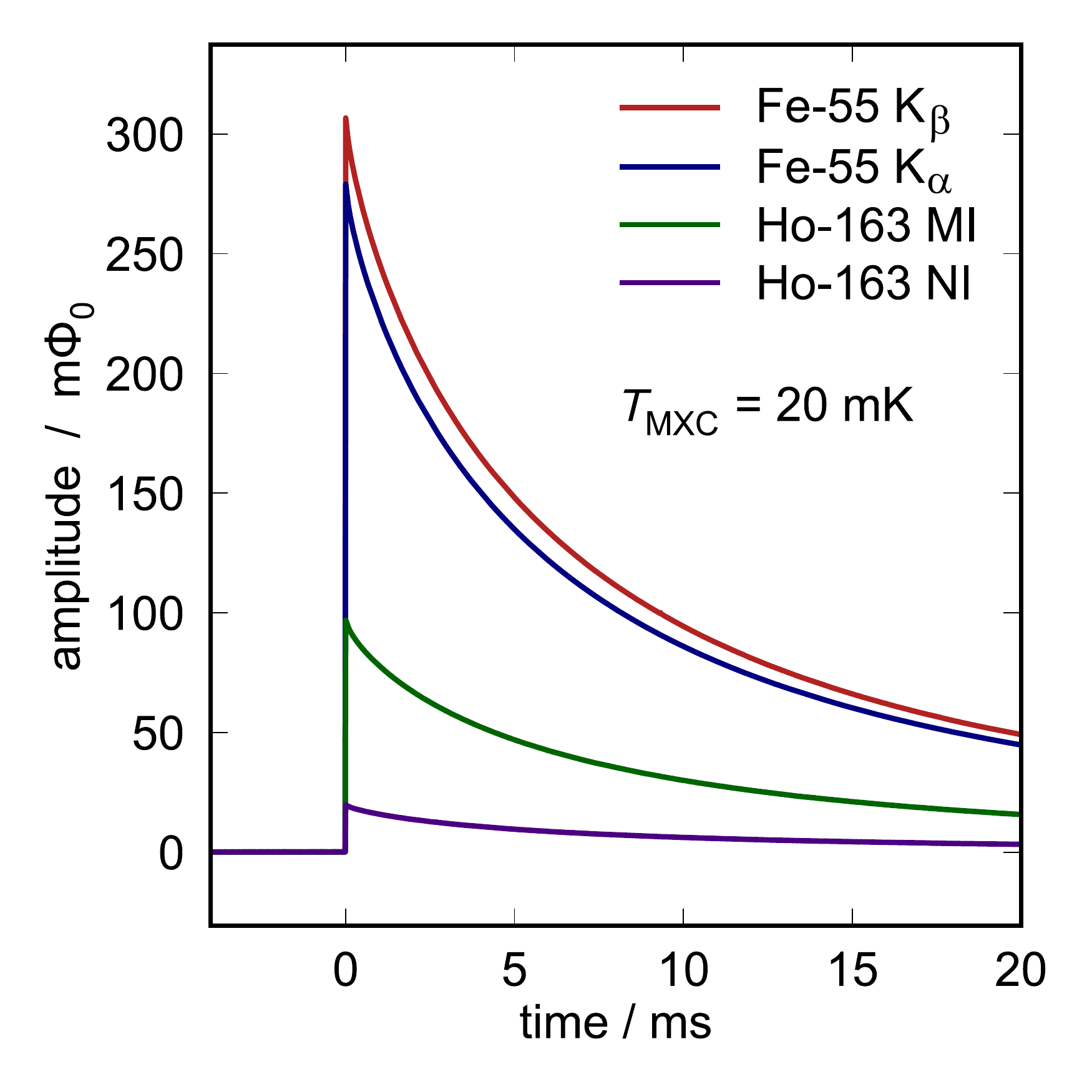}
        \caption{} \label{SUBFIG:shape_vs_energy_comparison_lin}
    \end{subfigure}
    \hfill
    \begin{subfigure}[b]{0.32\textwidth} 
        \centering
        \includegraphics[trim=12 0 12 0,clip,width=\linewidth]{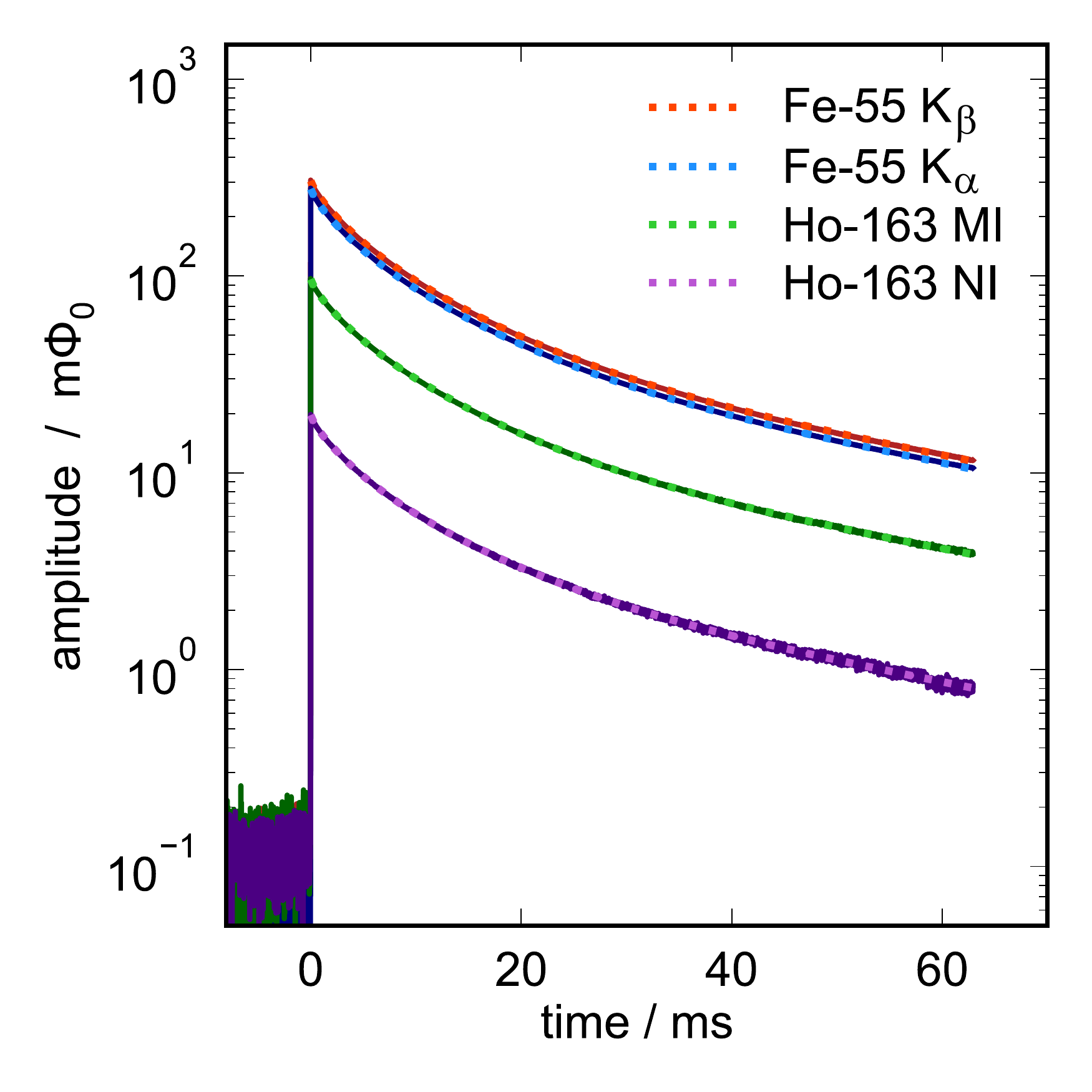}
        \caption{} \label{SUBFIG:shape_vs_energy_comparison_log}
    \end{subfigure}
    \hfill
    \begin{subfigure}[b]{0.32\textwidth} 
        \centering
        \includegraphics[trim=12 0 12 0,clip,width=\linewidth]{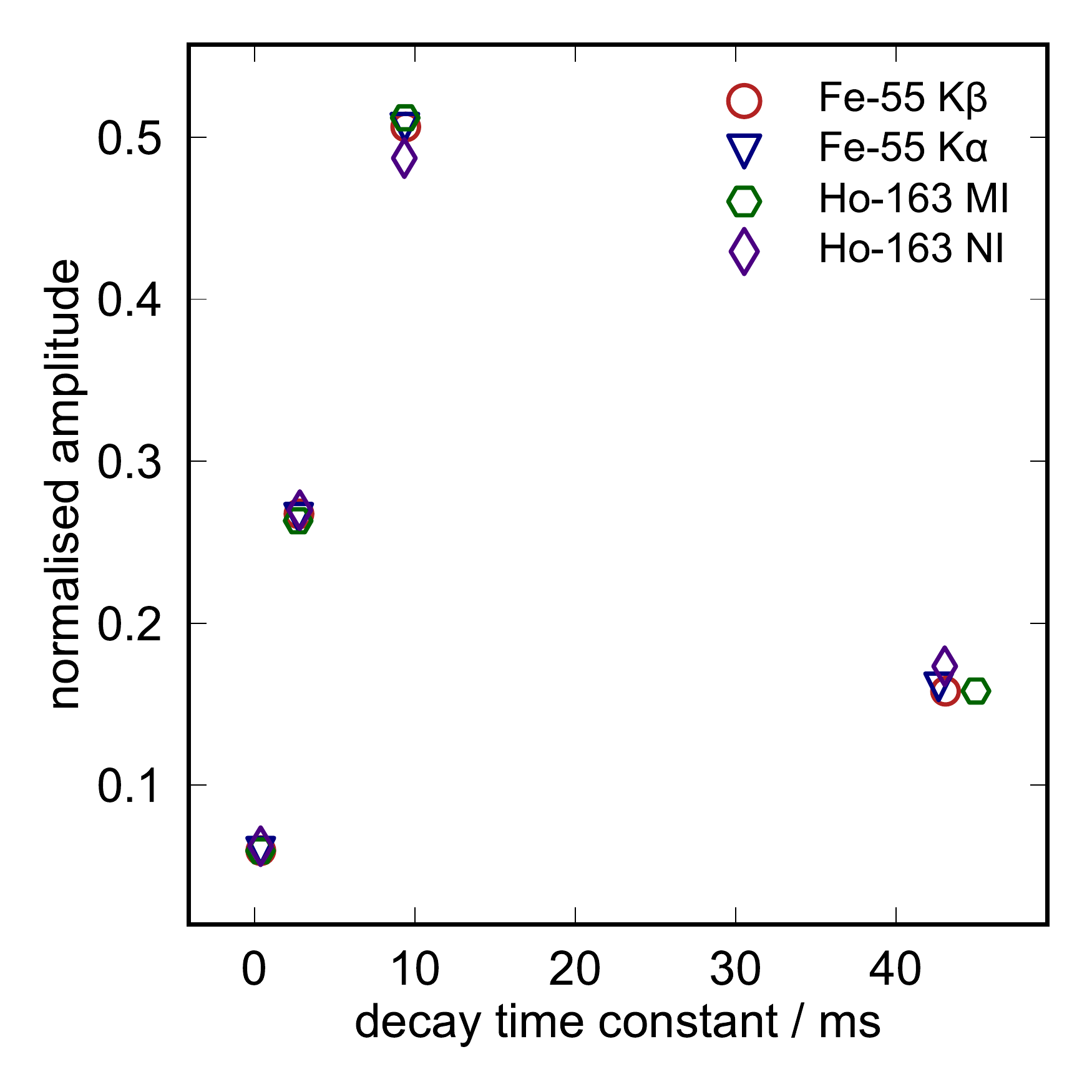}
        \caption{} \label{SUBFIG:fit_different_energies}
    \end{subfigure}

    \caption{Comparison between the average detector response for different energy input in linear scale \textbf{(a)} and logarithmic scale \textbf{(b)}. In (b) a multi-exponential fit is overlaid for each signal. The decay constants and the relative amplitudes returned by the fit are shown in \textbf{(c)}.}
    \label{FIG:shape_vs_energy_comparison}
\end{figure}

\paragraph{Detector response to different energy inputs}
In order to verify that the detector response is energy independent, signals from different spectral lines have been compared. Figure \ref{FIG:shape_vs_energy_comparison} shows the detector responses acquired at a MXC temperature of $30 \, \mathrm{mK}$ corresponding to the following energy inputs:

\begin{itemize}
    \item K$_\alpha$ photons with an energy of $5.89 \, \mathrm{keV}$ from a $^{55}$Fe source,
    \item K$_\beta$ photons with an energy of $6.49 \, \mathrm{keV}$ from a $^{55}$Fe source,
    \item MI electron capture radiation with an energy of $2.04 \, \mathrm{keV}$ from $^{163}$Ho source,
    \item NI electron capture radiation with an energy of $0.41 \, \mathrm{keV}$ from $^{163}$Ho source.
\end{itemize}

The detector signals have the same shape independently from the energy at all temperatures. 
Figure \ref{SUBFIG:shape_vs_energy_comparison_log} shows the signals in logarithmic scale with a fit performed with a multi-exponential function (equation \ref{EQ:multiexp}).
The decay constants $\tau_i$ and the normalised amplitudes returned by the fit shown in figure \ref{SUBFIG:shape_vs_energy_comparison_log} are summarised in the plot of figure \ref{SUBFIG:fit_different_energies}, showing that these parameters remain fairly unchanged for the different signals, proving that the shape of the decay does not depend on the energy deposition in the detector.

The rise time of the thermal pulses is ultimately limited by the coupling between the electronic system and the spin system of the sensor. 
However, in the performed measurements the main limitation is due to the read-out bandwidth, which depends on the FLL bandwidth. As a result, the typical rise time is electronically limited to be between $500 \, \mathrm{ns}$ and $1 \, \mathrm{\upmu s}$.

Overall, the energy-independent detector response allows for a reliable data analysis and the complex time profile of the signal decay does not affect the energy resolution for the expected rate of about $1 \, \mathrm{Bq}$, as planned for the ECHo-1k phase. 

\subsubsection*{$^{163}$Ho activity}
\label{SUBSUBSEC:ECHo-1k_activity}

The $^{163}$Ho activity per pixel has been determined for three implanted chips with $^{163}$Ho implanted in gold, silver and aluminium, respectively. Two independent methods have been used to estimate the $^{163}$Ho activity, which gave consistent results.

\begin{figure}[h!] 
    \centering
    \begin{subfigure}[b]{0.42\textwidth}
        \centering
        \includegraphics[trim=12 0 12 0,clip,width=\linewidth]{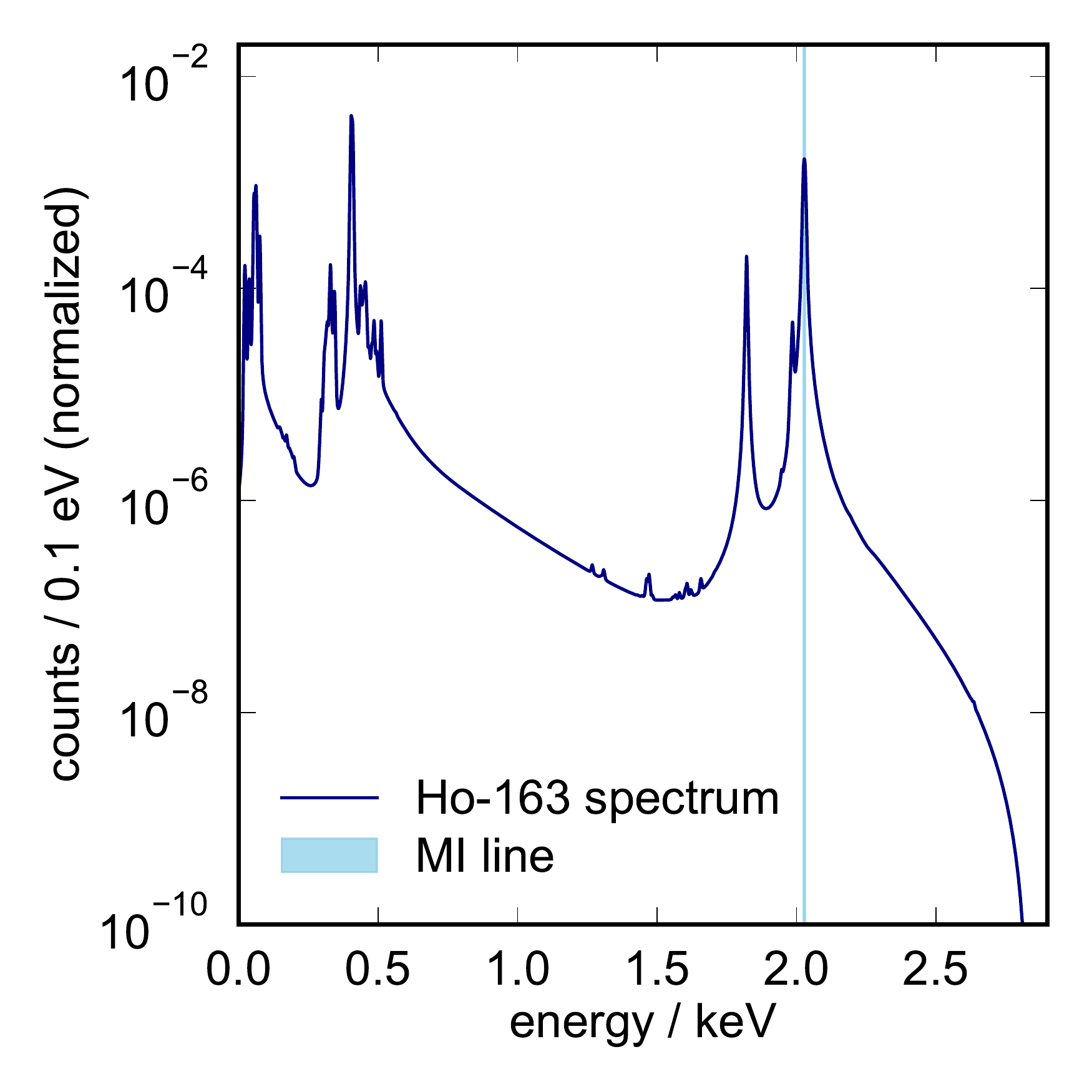}
        \caption{} \label{SUBFIG:spectrum_MI_line}
    \end{subfigure}
    \hfill
    \begin{subfigure}[b]{0.42\textwidth} 
        \centering
        \includegraphics[trim=12 0 12 0,clip,width=\linewidth]{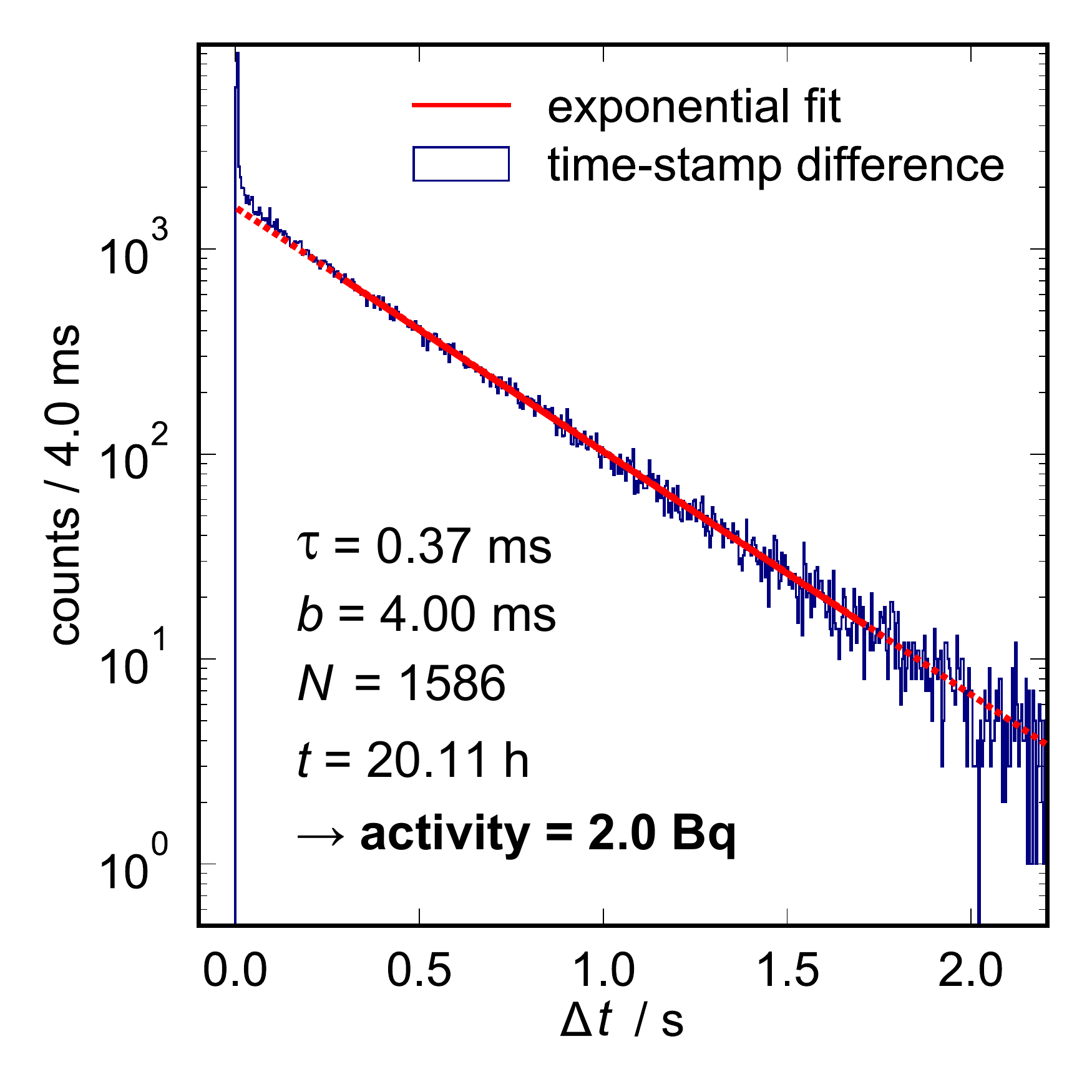}
        \caption{} \label{SUBFIG:activity_histo}
    \end{subfigure}

    \caption{\textbf{a)} Theoretical spectrum \cite{Brass_HoSpectrum_2020} with the energy range of the MI peak marked according to equation \ref{EQ:BR_MI}. \textbf{b)} Example of the time-stamp difference histogram in a semilogarithmic plot. The exponential fit is overlaid in red. The solid red line corresponds to the fit range, while the dashed red line shows the fitting function over the rest of the range. The extracted parameters are reported as well as the estimated activity.}

    \label{FIG:activity_methods}
\end{figure}

The first method is based on the integral of the $^{163}$Ho spectrum histogram in the energy range belonging to the MI peak. 
The activity cannot be extracted from the total number of recorded $^{163}$Ho events, since the energy threshold prevents to reliably detect events with energies below $\sim 30 \, \mathrm{eV}$ for all the implanted pixels.
\\
Using the recent theoretical description of the electron capture (EC) $^{163}$Ho spectrum presented in \cite{Brass_HoSpectrum_2020}, it is possible to derive the pseudo branching ratio of the MI line (i.e.~the percentage of events belonging to the MI line), $BR_\mathrm{MI}$:

\begin{equation} \label{EQ:BR_MI}
BR_\mathrm{MI} = \int_{0.99 \cdot E_{\mathrm{MI}}}^{1.01 \cdot E_{\mathrm{MI}}} S_\mathrm{N}(E) \, \mathrm{d}E = 0.216
\end{equation}

\noindent where $S_\mathrm{N}(E)$ is the normalised spectrum, as shown in figure \ref{SUBFIG:spectrum_MI_line}, and $E_\mathrm{MI} = 2.04 \, \mathrm{keV}$ is the energy of the MI line.
\\
The activity per pixel, $a$, can be derived as:

\begin{equation}
    a = \frac{N_{\mathrm{MI}}}{BR_{\mathrm{MI}}} \cdot \frac{1}{t}
\end{equation}

\noindent where $N_{\mathrm{MI}}$ is the number of events in the MI peak (i.e.~between $0.99 \cdot E_{\mathrm{MI}}$ and $1.01 \cdot E_{\mathrm{MI}}$) and $t$ is the acquisition time.
\\
The dominant error in this method is the systematic uncertainty of the branching ratio of the MI line, since it is based on a not yet perfect theoretical model. This uncertainty is estimated to be below 5 \%, based on the maximum deviation between the experimental spectrum and the theoretical model \cite{Brass_HoSpectrum_2020}.

The second method exploits the analysis of the $^{163}$Ho event rate. Each recorded detector signal is characterised by a time-stamp given by the ADC clock, which corresponds to the time when the pulse is triggered. The distribution of the time-stamp differences between two consecutive pulses $\Delta t$ in one detector pixel follows an exponential decay:

\begin{equation}
    \frac{\mathrm{d}N(\Delta t)}{\mathrm{d} (\Delta t)}  = \frac{N_\mathrm{0}}{b}  \cdot e^{- \Delta t / \tau}
\end{equation}
\label{EQ:TS_exp_decay}

\noindent where $N_\mathrm{0}$ is the number of counts in the bin where $\mathrm{\Delta} t = 0$, $\tau$ is the characteristic decay-time and $b$ is the bin width set at $b = 4 \, \mathrm{ms}$. \\
From a fit of the time-stamp difference histogram, the decay-time $\tau$ and $N_\mathrm{0}$ can be extracted. \\
The $^{163}$Ho activity $a$ for each pixel is then given by integrating the fit function and dividing by the acquisition time $t$:

\begin{equation}
    a = \frac{1}{t} \int_0^{\infty}  \frac{\mathrm{d}N(\Delta t)}{\mathrm{d} (\Delta t)} \mathrm{d}(\mathrm{\Delta}t)  = \frac{N_\mathrm{0} \cdot \tau}{b \cdot t} 
\end{equation}
\label{EQ:activity_from_fit}

Figure \ref{SUBFIG:activity_histo} shows an example of a time-stamp difference histogram for an ECHo-1k detector pixel with an exponential fit as it has been used to extract the activity. The spike present at very low $\Delta t$ is due to triggered noise, but it does not affect the activity estimation, since the fit window excludes the corresponding range.

The dominant uncertainties of this method are due to the possible presence of noise traces entering the time-stamp difference histogram and to the finite energy threshold that excludes a fraction of low-energy events from the histogram. The fit range is adapted to take into account only the central part of the histogram, which should not be relevantly affected by noise. An uncertainty on the final activity of about 5 \% has been estimated, comparing the activity values for the same detector pixel from different measurements based on this method.

\begin{figure}[h!] 
\begin{subfigure}{1.\textwidth}
  \centering
  \includegraphics[height=.14\textheight]{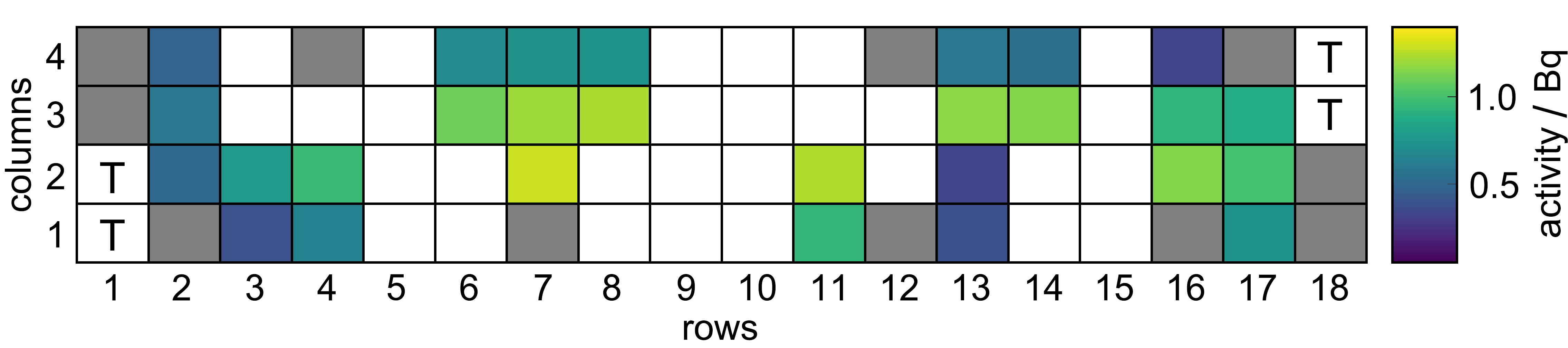} 
  \caption{ECHo-1k chip with gold as host-material}
  \label{SUBFIG:Au_colormap}
\end{subfigure}
\begin{subfigure}{1.\textwidth}
  \centering
  \includegraphics[height=.14\textheight]{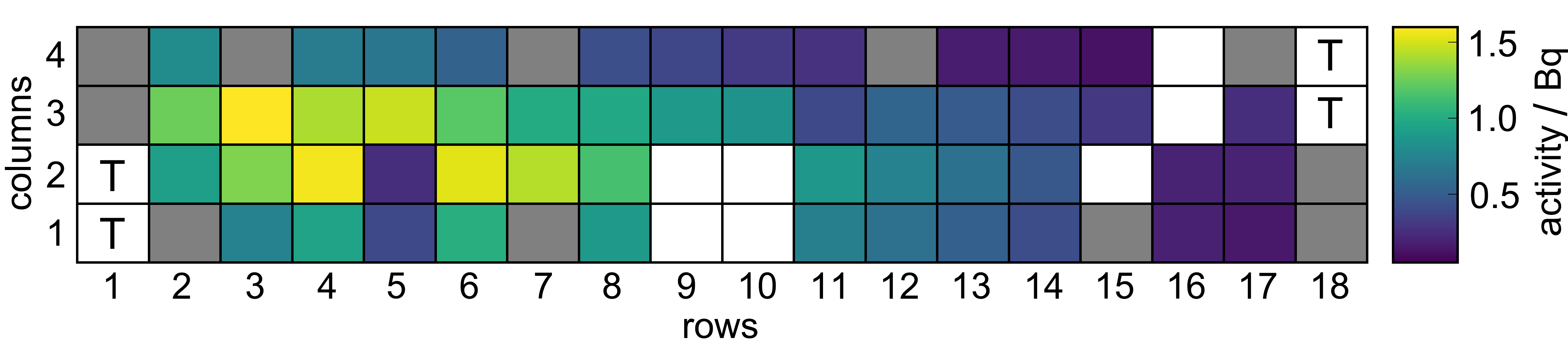}  
  \caption{ECHo-1k chip with silver as host-material}
  \label{SUBFIG:Ag_colormap}
\end{subfigure}
\begin{subfigure}{1.\textwidth}
  \centering
  \includegraphics[height=.14\textheight]{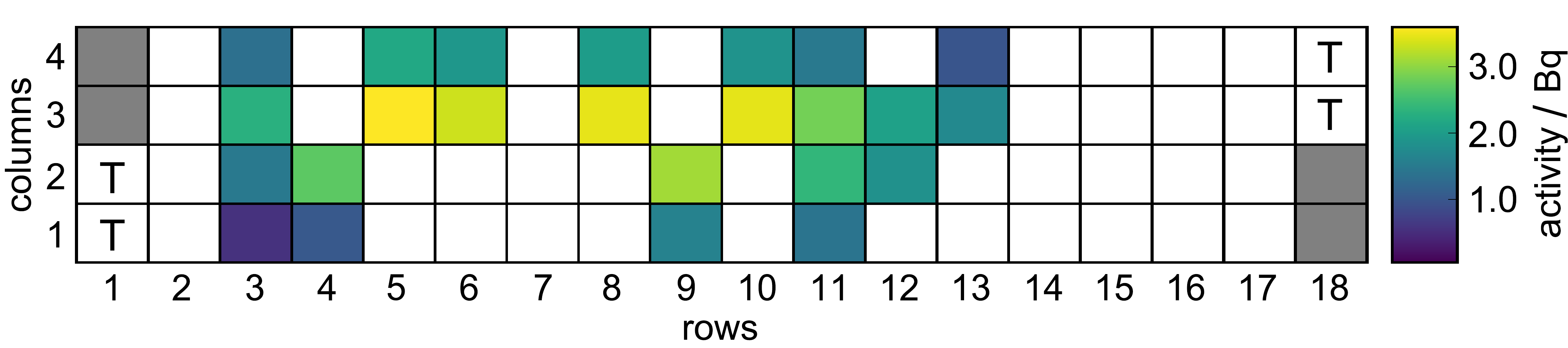}  
  \caption{ECHo-1k chip with aluminium as host-material}
  \label{SUBFIG:Al_colormap}
\end{subfigure}
\caption{Activity maps of the characterised ECHo-1k chips. The colour code refers to the activity. The pixels marked in grey are non-implanted. The pixels in white have not been measured. The temperature pixels are marked with the letter T.}
\label{FIG:colormaps}
\end{figure}

The resulting activity values per pixel of the three ECHo-1k chips are shown in the colour maps of figure \ref{FIG:colormaps}. Not all the MMC pixels could be characterised due to missing connections of the corresponding read-out channels.
The average activity and the corresponding standard deviation of each ECHo-1k chip are reported in table \ref{TAB:average_std_activity}.

\begin{table}[h!]
\centering
\begin{tabular}{ |c|c|c| } 
 \hline
 Implantation material & Average activity [Bq] & Standard deviation [Bq] \\ 
 \hline
 Gold & 0.81 & 0.30 \\ 
 Silver & 0.71 & 0.44 \\ 
 Aluminium & 2.12 & 0.84 \\
 \hline
\end{tabular}
\caption{Summary of the obtained average $^{163}$Ho activity values and corresponding standard deviations for the three ECHo-1k chips with different $^{163}$Ho host material.}
\label{TAB:average_std_activity}
\end{table}

The considerably larger average activity for the ECHo-1k chip where $^{163}$Ho has been implanted in aluminium with respect to the other chips matches the expectations. The larger implantation depth in aluminium predicted by simulations \cite{implantation_sim} limits the back-sputtering of previously implanted $^{163}$Ho ions and therefore allows to more easily reach higher activity values.

In all three chips a trend of higher activities in the two central rows 2 and 3 as compared to the outer rows 1 or 4 is observed, which confirms the overall adjustment of the extended $^{163}$Ho ion beam spot of typically $0.6 \, \mathrm{mm}$ diameter. When this beam is scanned over the active area of the 72 MMC pixels (in total $5 \, \mathrm{mm}$ in width and $1.2 \, \mathrm{mm}$ in height), a slight overloading by about a factor of 2 of the central area is unavoidable and well acceptable. 
In addition, the chip with silver as host material (figure \ref{SUBFIG:Ag_colormap}) shows a clear asymmetry with higher activities on the left than on the right hand side. This is attributed to imperfect horizontal ion beam steering during that particular implantation. 
Additionally, this chip with silver as host material also exhibits a substantial activity variation in the neighbouring pixels in row 5. This might be due to a not perfect development of the photoresist at this position. Photoresist residuals on the implantation area might thus have caused only a fraction of the incoming $^{163}$Ho ions to actually be implanted.

In order to gain more precise control over the implanted activity and to reach the exact planned activity per pixel, further optimisation and automatised control of the ion beam steering during implantation at RISIKO will be developed.
The goal for the next phase of the ECHo experiment is to implant $10 \mathrm{Bq}$ per pixel with optimum homogeneity, i.e.~a variation of well below a factor of 2.

\subsubsection*{Energy resolution}

The two ECHo-1k chips with $^{163}$Ho implanted in gold and silver, respectively, have been selected to run the high-statistics measurement as conclusion of the ECHo-1k phase. \\
The detector chips have been operated at about $15 \, \mathrm{mK}$ and the detector performances could be evaluated during the run, also in terms of energy resolution per pixel.

\begin{figure}[h!] 
	\centering
	\includegraphics[width=0.75\textwidth]{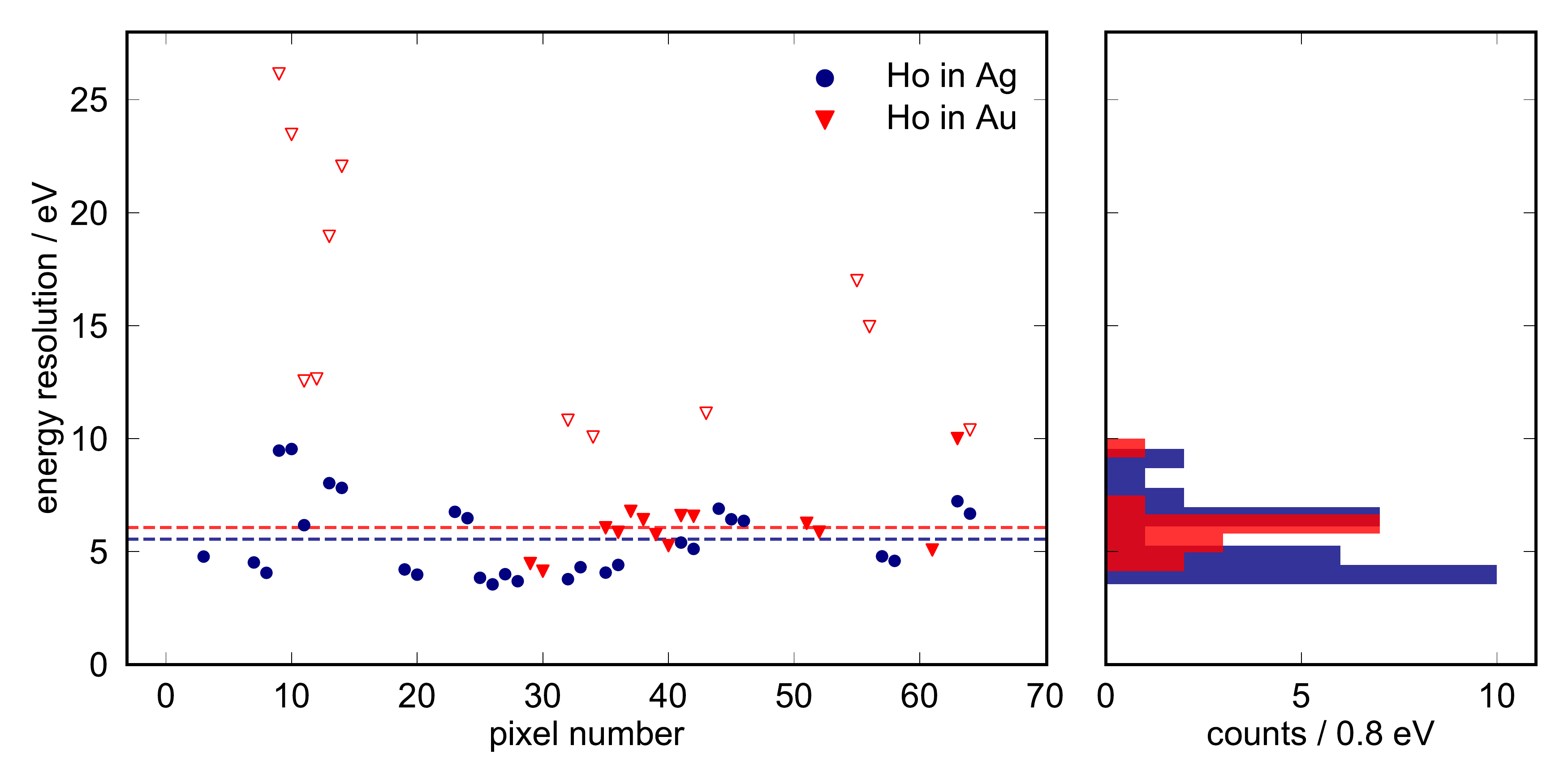}
	\caption{Energy resolution $\mathrm{\Delta}E_\mathrm{FWHM}$ per pixel of the two ECHo-1k chips under investigation. In red (blue) the data for the chip with gold (silver) as host material are shown. The unfilled markers correspond to the pixels with insufficient energy resolution. The average energy resolution values for the two set-ups are shown with dashed lines of the corresponding colour.} 
	\label{FIG:en_res}
\end{figure}

The energy resolution can be extracted from a fit of the NI line (corresponding to the capture of 4s electrons) at the energy $E_{\mathrm{NI}} = 411 \, \mathrm{eV}$ \cite{Ranitzsch_JLTP}. The fit function is a convolution of a Lorentzian distribution, which is close to the shape of the NI peak, and a Gaussian curve, which takes into account the detector response.
The width parameter of the Lorentzian is fixed to $5.3 \, \mathrm{eV}$ \cite{Ho_widths}. The standard deviation of the Gaussian is returned by the fit and from that the energy resolution can be determined as full width at half maximum.
Figure \ref{FIG:en_res} shows the energy resolution per pixel of the two ECHo-1k chips eva\-lu\-ated for the first data-set acquired for the high-statistics measurement.
The data points shown with unfilled markers correspond to the detector pixels with an energy resolution worse than the benchmark for the ECHo-1k phase, i.e.~$10 \, \mathrm{eV}$ FWHM. This not optimal performance could be improved by optimising the shielding against electromagnetic noise. These 12 pixels have been excluded from the histogram in figure \ref{FIG:en_res} and from the following averaging procedure. 

The average energy resolution for the chip with gold as host material is $\langle \mathrm{\Delta}E_\mathrm{FWHM} \rangle = 6.07 \, \mathrm{eV}$ with a standard deviation of $\sigma = 1.33 \, \mathrm{eV}$. 
The average energy resolution for the chip with silver as host material is $\langle \mathrm{\Delta}E_\mathrm{FWHM} \rangle = 5.5 \, \mathrm{eV}$ with a standard deviation of $\sigma = 1.7 \, \mathrm{eV}$. 
The standard deviations of just about 20 or 30 \% for gold or silver, respectively, show that the energy resolution is fairly similar for all the pixels. This homogeneity ensures a good energy calibration for each single data sample, which in turn guarantees a reliable sum of the corresponding single spectra. Therefore, is it possible to properly merge all the single $^{163}$Ho spectra in order to build the final high-statistic spectrum. 

The estimation of the energy resolution per pixel is the first quality check step in the analysis routine that is implemented for each data-set. In fact, different environmental conditions or noise sources can alter the detector pixel performances during the measurement.

From the theoretical calculation of the detector performances, assuming an implanted $^{163}$Ho activity of $a = 1.0 \, \mathrm{Bq}$ and the measured read-out noise levels, the expected energy resolution is $\Delta E_{\mathrm{FWHM}} \approx 5.15 \, \mathrm{eV}$ at $20 \, \mathrm{mK}$. This result is fully confirming the experimental observations.

\section{Summary and conclusion}

For the first phase of the ECHo experiment, ECHo-1k, a dedicated 72-pixels MMC chip has been developed. The ECHo-1k chip has been successfully produced employing photo-lithographic microfabrication techniques. The preliminary characterisation measurements at room temperature as well as the functionality checks at $4.2 \, \mathrm{K}$ gave very promising results for all the tested chips, which have been confirmed by the excellent performance observed during the operation at millikelvin temperature.
A customised ion-implantation procedure has been optimised in order to embed $^{163}$Ho in the detector absorbers, ensuring high purity of the source and a complete 4$\pi$ enclosure. For that, specific post-implantation photo-lithographic steps have been developed. 
Two of the implanted chips, one with silver and the other one with gold as $^{163}$Ho host material, have been operated at millikelvin temperature in a dilution refrigerator and they have been fully characterised in terms of pulse shape, activity and energy resolution. The detector pulse shape has been analysed for MMC pixels with and without implanted $^{163}$Ho at different temperatures - in a range between $20 \, \mathrm{mK}$ and $120 \, \mathrm{mK}$. 

The $^{163}$Ho activity per pixel has been determined and the energy resolution as FWHM has been calculated from a fit of the NI line. An average activity of $0.81 \, \mathrm{Bq}$ ($0.71 \, \mathrm{Bq}$) and an average energy resolution FWHM of $6.1 \, \mathrm{eV}$ ($5.6 \, \mathrm{eV}$) have been calculated for the chip with gold (silver) as host material, matching the requirements for the first phase of the ECHo experiment. These two ECHo-1k chips have been used for a high-statistics acquisition of more than $10^8$ $^{163}$Ho events. This level of statistics will allow to derive a new limit on the effective electron neutrino mass below $20 \, \mathrm{eV}$ 90\% C.L..

\section*{Acknowledgements}

Part of this research was performed in the framework of the DFG Research Unit FOR2202 “Neutrino Mass Determination by Electron Capture in 163Ho, ECHo” (funding under DU 1334/1-1 and DU 1334/1-2, EN 299/7-1 and EN 299/7-2, EN 299/8-1, GA 2219/2-1 and GA 2219/2- 2). F. Mantegazzini, A. Barth, C. Velte and M. Wegner acknowledge support by the Research Training Group HighRR (GRK 2058) funded through the Deutsche Forschungsgemeinschaft, DFG. Furthermore, we thank the cleanroom team at the Kirchhoff-Institute for Physics for technical support during device fabrication.

\bibliographystyle{elsarticle-num}
\bibliography{bib_MMCs,bib_ECHo,bib_misc}

\end{document}